\documentclass[]{aa}

\usepackage{graphicx}
\usepackage{natbib}
\title{The Bullet cluster at its best: weighing stars, gas and dark matter}

\begin{document}

\author{D. Paraficz\inst{\ref{EPFL}, \ref{Marseille}}, J.-P. Kneib\inst{\ref{EPFL}}, J. Richard\inst{\ref{Lyon}}, A. Morandi\inst{\ref{USA}}, M. Limousin\inst{\ref{Marseille}}, E. Jullo\inst{\ref{Marseille}}, Johany Martinez\inst{\ref{Lyon}}}
\institute{Laboratoire d'Astrophysique
Ecole Polytechnique F\'ed\'erale de Lausanne (EPFL)
Observatoire de Sauverny
CH-1290 Versoix \label{EPFL}
\and
Aix Marseille Universit\'e, CNRS, LAM (Laboratoire d'Astrophysique de Marseille) UMR 7326, 13388, Marseille, France \label{Marseille}
\and
Univ Lyon, Univ Lyon1, Ens de Lyon, CNRS, Centre de Recherche Astrophysique de Lyon UMR5574, F-69230, Saint-Genis-Laval, France \label{Lyon}
\and
Physics Department, University of Alabama in Huntsville, Huntsville, AL 35899, USA \label{USA}
}
\date{\today}

\titlerunning{The Bullet cluster revisited}
\authorrunning{D. Paraficz}

%==============================================================================================================================

\abstract
{}
{We present a new  strong lensing mass reconstruction of the ``Bullet cluster'' (1E 0657-56) at z=0.296,  based on  WFC3 and ACS HST imaging and VLT/FORS2 spectroscopy.
The strong lensing constraints underwent substantial revision compared to previously published analysis, there are now 14 (six new and eight previously known)  multiply-imaged systems, of which three have spectroscopically confirmed redshifts  (including one newly measured from this work).}
{The reconstructed mass distribution explicitly included the combination of three mass components:  $i)$ the intra-cluster gas mass derived from X-ray observation, 
$ii)$ the cluster galaxies modeled  by their fundamental plane scaling relations and  
$iii)$ dark matter.}
 {The model that includes the intra-cluster gas is the one with the best Bayesian evidence. This model
has a total RMS value of $0.158\arcsec$ 
between the predicted and measured image positions  for the 14 multiple images considered. The proximity of the total RMS  to resolution of HST/WFC3 and ACS (0.07-0.15" FWHM) demonstrates the excellent precision  of our mass model.
The derived mass model confirms the  spatial offset between the X-ray gas and dark matter peaks.
The fraction of the galaxy halos mass to total mass  is found to be $f_{s}=11\pm5$\% for a total mass  of $2.5\pm0.1 \times 10^{14} M_{\odot}$ within a 250 kpc radial aperture. }
{}
%==============================================================================================================================

\maketitle

%==============================================================================================================================

\section{Introduction}
    \begin{figure*}
\centering
\includegraphics[width=0.95\textwidth]{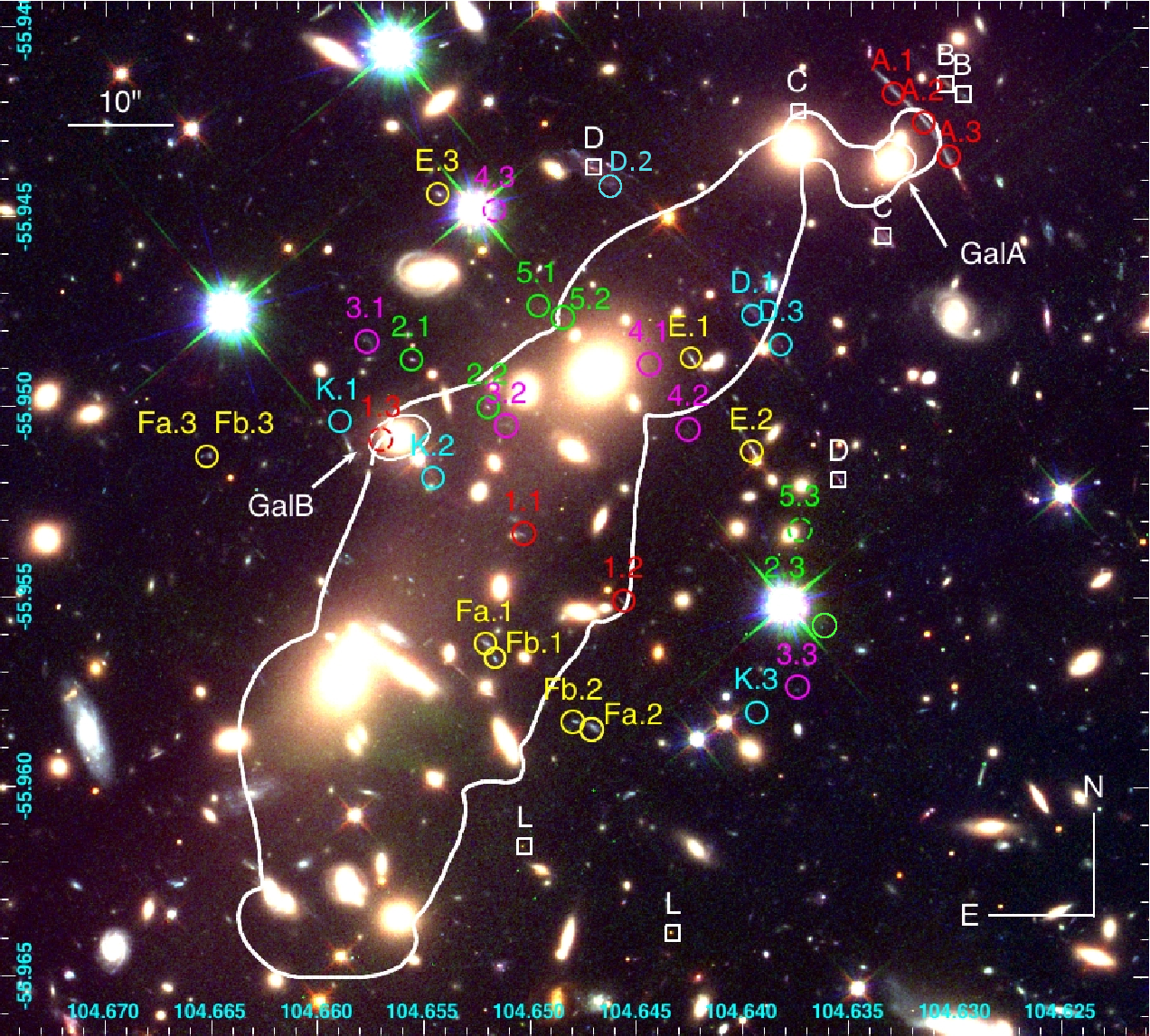}
\caption{Color HST image  of the main cluster component of 1E 0657-56 (blue--F606W, green--F814W, red--F160W).  Multiple images considered in this work are marked with color circles (dashed line circles mark the predicted but not confirmed positions of counter images), the spectroscopically confirmed multiply-imaged systems are  system A \citep{Mehlert:2001} and K \citep{Gonzalez:2010}. White squares are referring to \citep{Bradac:2009} systems, which we have revised and  we did not include in our modeling.  
The new identification of system A is shown in red and the new identification of system D is shown in cyan (see also Figure~\ref{DMulti} ). System D is a multiply-imaged candidate and due to extended morphology is not a part of model constraints. The white line represents a critical line corresponding to $z=3.24$. }
\label{MainClump}
\end{figure*}

    \begin{figure*}
\centering
\includegraphics[width=0.95\textwidth]{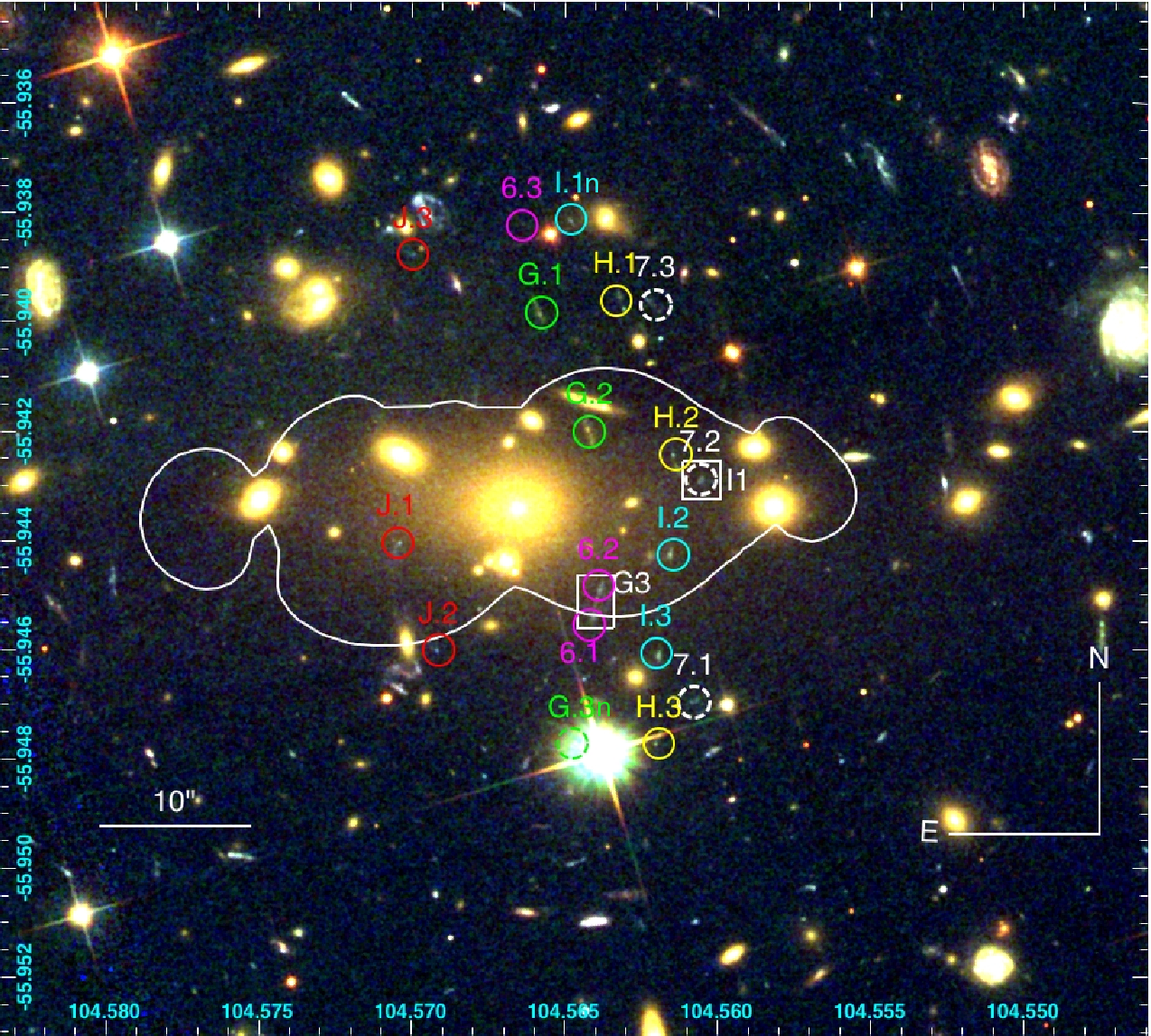}
\caption{Color ACS/HST image of the sub cluster component of 1E 0657-56 (blue--F435W, green--F606W, red--F814W). Multiple images considered in this work are marked with color circles 
 (dashed line circles mark the predicted but not confirmed positions of counter images).
 In this work,  we spectroscopically measured a redshift of $z=2.99$ for the multiply-imaged system H (see section 2.3). System 7 (white dashed circle)
is a multiply-imaged candidate and is not a part of model constraints due to large color uncertainties.
The white line represents the critical line at redshift $z=2.99$.}
\label{SubClump}
\end{figure*}

The massive galaxy cluster, 1E 0657-56, discovered by \cite{Tucker:1998} consists of  two colliding galaxy clusters at z = 0.296.  In this  distinct merging system, a sub-cluster, the ``bullet'' has collided with the main cluster, approximately in the plane of the sky  \citep{Barrena:2002}. The bullet-like sub-cluster  has produced strong bow shock in the intra-cluster gas during the collision and consequently  the collision stripped the gas from the cluster potential \citep{Markevitch:2002}.
The offset between the two baryonic components (gas and galaxies) gave a remarkable possibility for the  indirect measurements of the total mass distribution using gravitational lensing studies \citep{Mehlert:2001, Clowe:2004, Bradac:2006, Bradac:2009}, which unambiguously demonstrated that dark matter (DM) traces the colisionless galaxies and not the X-ray gas.
The study of the lensing mass distribution of the Bullet cluster remains a powerful evidence of the DM existence that severely challenges theories of modified gravity such as MOND and TeVeS \citep{Milgrom:1983, Bekenstein:2004}. It also gives
upper limits on the DM self interaction cross section  \citep{Randall:2008,Markevitch:2004}  and lower limit on the possible radiative decay of  DM  \citep{Boyarsky:2008}.

Since the first lensing mass measurement of the Bullet cluster  considerable effort has been put in constraining its mass distribution. Nevertheless, significant discrepancies exist, for example, the masses  at R $<$ 200 kpc derived  from the \cite{Clowe:2004} and \cite{Bradac:2006} differ by a factor of two.  
Although, this is likely due to degeneracies of lens modeling between strong and weak lensing mass estimates, the \cite{Bradac:2006} mass measurement uncertainty is still as high as 14\% over the full ACS field.
Indeed, the complexity of the Bullet cluster and the limited number of multiple images makes its strong lens modeling exceptionally challenging.

 As witnessed in other clusters  \citep[{e.g.,} Abell 1689, 1703, 2218, see][]{Richard:2010}, the accuracy of the lensing mass map is strongly dependent on the correct identification and on the number of multiply-imaged systems used to constrain it. This accuracy  has a further  impact on the measurements of magnification of high redshift galaxies  \citep{Bradac:2009, Hall:2012} and the luminosity function estimations. Hence, to construct a robust mass model of an accurate gravitational telescope  many spectroscopically confirmed multiply-imaged systems are needed. Therefore, The Bullet cluster does not compete  with the   Frontier Field clusters as gravitational telescope  \citep{Jauzac:2015, Richard:2014}, nevertheless the  methods developed in this work, could be implemented also to other galaxy  clusters.

In this work we present an improved high-resolution strong lensing mass model of the Bullet cluster (the combination of weak lensing and our newly reconstructed strong lensing model will be published in our next paper)  based on the identification of new multiply-imaged systems. These results are based on Advanced Camera for Surveys (ACS)  and  Wide Field Camera 3 (WFC3) new  images as well as on new spectroscopic redshift determination of multiply-imaged  systems, one taken from the literature and one obtained through VLT/FORS2 observations.
Furthermore, our mass  reconstruction includes  the novel combination of the following mass components:  $i)$ the intra-cluster gas mass derived from X-ray observation, $ii)$ the cluster galaxies modeled  by their fundamental plane  scaling relations and  $iii)$ dark matter.
The gas mass component is  distinctive in the Bullet cluster since the gas is spatially shifted from the main mass component of the cluster \citep [$\sim 47\arcsec$] []{Clowe:2006} and has been thoroughly studied.
Nevertheless, until now the  X-ray data of a gas component of this cluster has not been independently taken into account
in any of the  lens modeling. Enriching our modeling technique and improving the lensing constraints allowed us to create a more accurate mass model with significantly smaller systematic uncertainties.

This paper is structured as follows. Section 2 describes the data and its reduction procedures. Section 3 describes the previous achievements in the field and presents our multiply-imaged systems.
Section 4 presents the method of mass reconstruction of the Bullet cluster, describes newly implemented mass modeling improvements (new scaling relations and X-rays mass map).  Our  conclusions are summarized in Section 5.
  
Throughout the paper, we assume a $\Lambda$ cold dark matter ($\Lambda$CDM) cosmology with $\Omega_m=0.3$, $\Omega_\Lambda=0.7$ and $h_{100} = 0.7$. At the cluster redshift $z=0.296$, 1'' corresponds to 4.413 kpc.

The reference center of our analysis is fixed at the BCG 1 center: $\alpha=$104.6588589
 $\delta=$-55.9571863 (J2000.0). Magnitudes are given in the AB system.   Unless stated otherwise, all uncertainties and upper and lower limits
are given and/or plotted at $1\sigma$ confidence level.

%==============================================================================================================================

\section{Observations and data reduction}
%==============================================================================================================================

\subsection{Hubble imaging}

The first round of observation  of the cluster 1E 0657-56 with the Hubble Space Telescope (HST) was carried out between 2004 and 2006 using the ACS camera (HST programs 10200 and 10863, PI: Jones \& Gonzalez) at two side-by-side positions covering the main cluster and the Western sub-cluster. The main cluster was observed in F606W, F775W and F850LP bands 
(hereafter V, i, z ), and the sub-cluster in F435W, F606W, F814W (hereafter B, V, I) (see Figure~\ref{MainClump} and \ref{SubClump}). The F606W band covering both 
components is used for a uniform view of the cluster, as well as detections in the photometric catalog. Details on the exposure times and quality of these data have been presented in \cite{Bradac:2009}.

In addition, HST/WFC3 imaging has been performed in F110W and F160W bands, with two largely overlapping 
positions centered on the main cluster (PID: 11099; PI: Brada{\v c}), and  with a single pointing at the center of the main cluster (PID:11591; PI: Kneib). The total 
exposure time in F110W/F160W was 6529/7029 secs for the first program, and 3211/2811 secs for the second program, with 9740/9840 secs in the overlapping region.  The magnitude limits at $3-\sigma$ measured using $0.25\arcsec$  radius aperture the deepest data reach  28.76 and 28.15 in F110W and F160W band, respectively.

Finally, in February 2011, the main cluster of 1E 0657-56 was observed with the ACS camera using the F814W filter as part of the program 11591. The total exposure time was 4480 secs (2 orbits). 

Both ACS and WFC3 data have been aligned using the multidrizzle \citep{Koekemoer:2002} software, including some 
relative shifts measured with \textsc{IRAF}\footnote{IRAF is distributed by the National Optical Astronomy Observatory, which is operated by the Association of Universities for Research in Astronomy (AURA) under cooperative agreement with the National Science Foundation.} for datasets taken at different epochs. The F606W image was used for overall 
alignment of the different bands, and the USNO B1.0 catalog provided absolute astrometric calibration. 

We use the double-image mode (with F606W being a detection image) of the \textsc{SExtractor} package \citep{Bertin:1996} to detect objects and compute magnitudes	within a $0.5\arcsec$  diameter aperture (ACS images). 
All of our imaging data (optical/ACS and near-IR/WFC3) are PSF-matched to the WFC3/IR F160W imaging data before making color measurements. We measure isophotal magnitudes to produce accurate colors and photometric redshifts.

The half-light radius $R_{\rm eff}$  used in the fundamental plane galaxy scaling was measured  using the \textsc{Galapagos}  on F606W image.

\subsection{Cluster member identification}
%==============================================================================================================================

\begin{figure}
\centering
\includegraphics[scale=0.35]{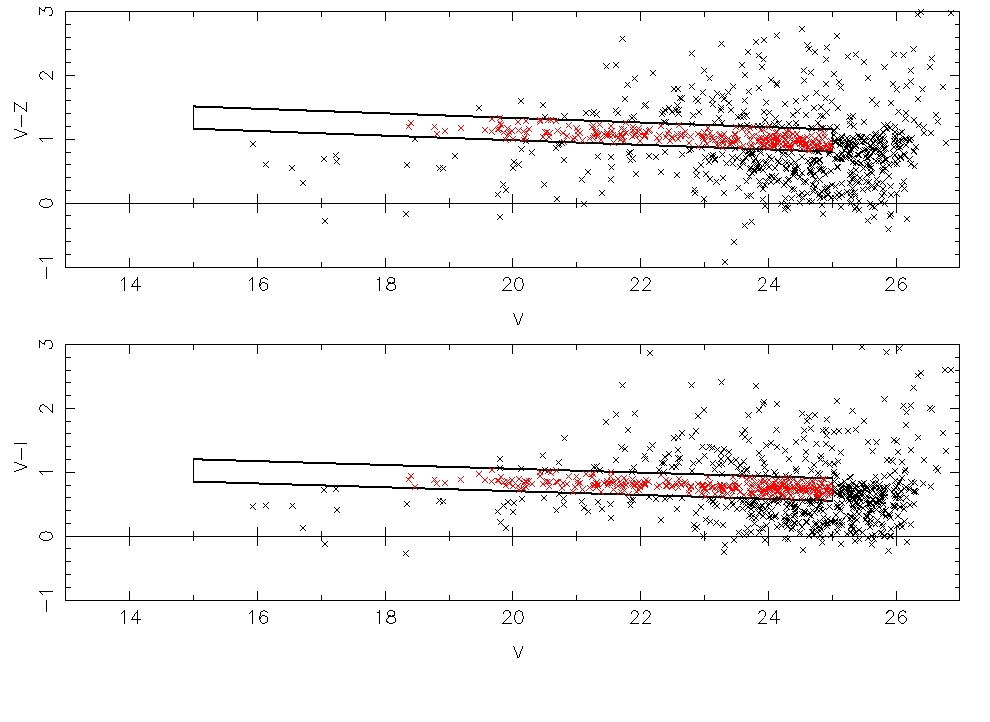}
\caption{Color-magnitude diagrams and the selection of cluster member galaxies. The red sequence selection is shown in the black boxes: all galaxies in this box  are
considered to be  cluster galaxies. [(V-I) vs. V] or [(V-z) vs. V] color-magnitude diagrams correspond to the main and sub-cluster components, respectively.}
\label{ColorDiagram}
\end{figure}

Cluster galaxies were identified based on the ACS data using the characteristic cluster red-sequences identified using the [(V-I) vs. V] or [(V-z) vs. V] color-magnitude diagrams (see Figure~\ref{ColorDiagram}), for the main and sub-cluster components, respectively.  The galaxies lying in one of the red sequences were assumed to be cluster members. 
 In order to save computing time we have included in the lens modeling only the 100 brightest cluster galaxies ($V<25.2$), that roughly corresponds to lensing deflection larger than $\sim 0.1\arcsec$).

\subsection{VLT/FORS2 spectroscopy}

We have used the FOcal Reducer and low dispersion Spectrograph
(FORS2, \cite{Appenzeller:1998}) at the Very Large Telescope to
measure the spectroscopic redshift of multiply-imaged systems. MXU masks with
 $1\arcsec$-wide slits were designed to cover most of the multiple images identified
in the Bullet cluster. Observations were obtained on the 3 nights of February 15-17th 2010,
with a total of 9.9 ksecs split into 900 seconds exposures.
The G300V grism and the GG435 order-sorting filter were used to provide a good coverage of
the reddest wavelengths ($4450<\lambda<8650$ \AA) a dispersion of 2.69 \AA{}   per pixel and a
resolution R=$\lambda/\Delta\lambda\sim$200 at the central wavelength 5900 \AA. Standard stars
were observed during the same nights, and the data reduction was performed with a combination
of the \textsc{esorex} package and standard \textsc{IRAF} routines to improve the sky subtraction and wavelength
calibration of specific slits.

\begin{figure}
\centering
\includegraphics[scale=0.3]{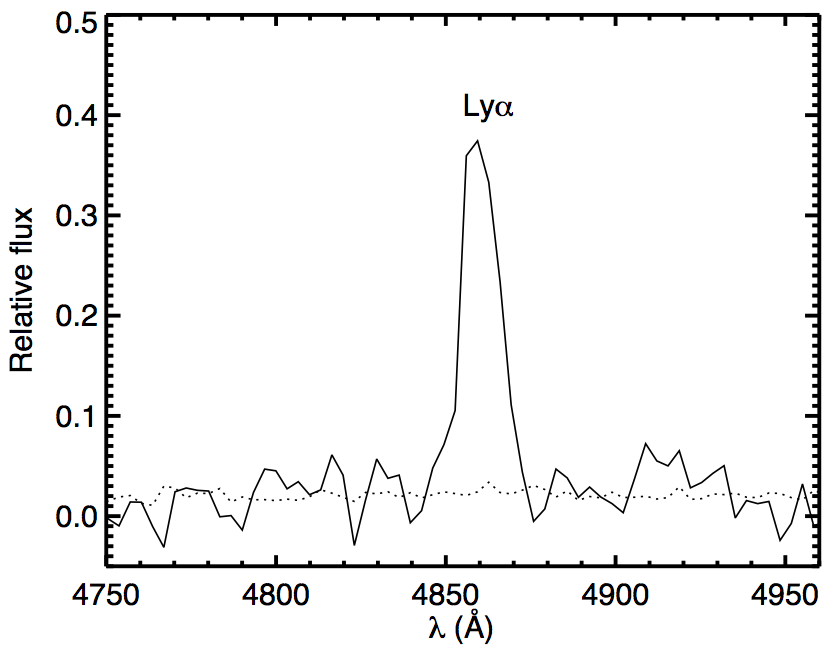}
\caption{ System H shows a strong emission line at 4851\AA{}  which we interpret as
Lyman-$\alpha$ at $z=2.99$ from the lensing configuration and the lack of
additional emission lines. The dotted line is the sky noise spectrum (no sky emission line are present in this wavelength range). The other likely alternative is $[{\rm O II}]$, observed at 4851\AA{} would give a redshift of 0.301, similar to the cluster redshift, thus the alternative can be excluded.}
\label{Hspec}
\end{figure}

Within these shallow spectroscopic data we only manage to measure the redshift for two systems.
We confirm the z=3.24 spectroscopic redshift of system A,
previously found by \cite{Mehlert:2001}. We also measured  the redshift of system H, one
of the multiply-imaged systems identified in the sub-cluster. This source
shows a strong emission line at 4851\AA{}  which we interpret as
Lyman-$\alpha$ at $z=2.99$ from the lensing configuration and the lack of
additional emission lines.  The other likely alternative is $[{\rm O II}]$, observed at 4851\AA{} would give a redshift of 0.301, similar to the cluster redshift, thus the alternative can be excluded  (see Figure~\ref{Hspec}).

%--------
\section{Multiple-image identification}

\subsection{Previous work}

The first lens model of the Bullet cluster, derived by \cite{Mehlert:2001} was based on 3 modeled DM clumps  as SIS and SIE and 150 cluster members modeled using the Faber-Jackson relation 
\citep{Faber:1976}. As lensing constraints \cite{Mehlert:2001} used 6 multiply-imaged systems (labeled A to F selected from deep BgRI VLT/FORS images). For one of the multiple systems (the giant arc) they measured a spectroscopic redshift of $z=3.24$.  

\cite{Bradac:2006,Bradac:2009} applied a  grid based mass reconstruction method based on strong and weak gravitational lensing (their weak lensing signal was taken from \cite{Clowe:2004, Clowe:2006}).
Using deep, high-resolution optical data from 3  ACS bands (F435W, F606W, and F814W),  BVR data from Magellan and I-band from VLT/FORS \citep{Clowe:2004} they confirmed (based on photometry and morphology) six multiply-imaged  systems  as discovered by 
\cite{Mehlert:2001} (labeled A--F) and also identified 4 new additional systems (G--J) in the sub-cluster region, where none were previously known. 
The combined  mass reconstruction of \cite{Bradac:2006} provided a high-resolution, absolutely calibrated mass map, with a projected, enclosed mass $M_{>250 \rm kpc} =2.8\pm0.2 \times 10^{14} M_{\odot}$ around the main cluster and $M_{>250 \rm kpc}=2.3\pm0.2 \times10^{14} M_{\odot}$ around the sub-cluster.

%------
\subsection{Critical evaluation of previously identified systems and new identification}

\begin{figure*}
\centering
\includegraphics[width=0.95\textwidth]{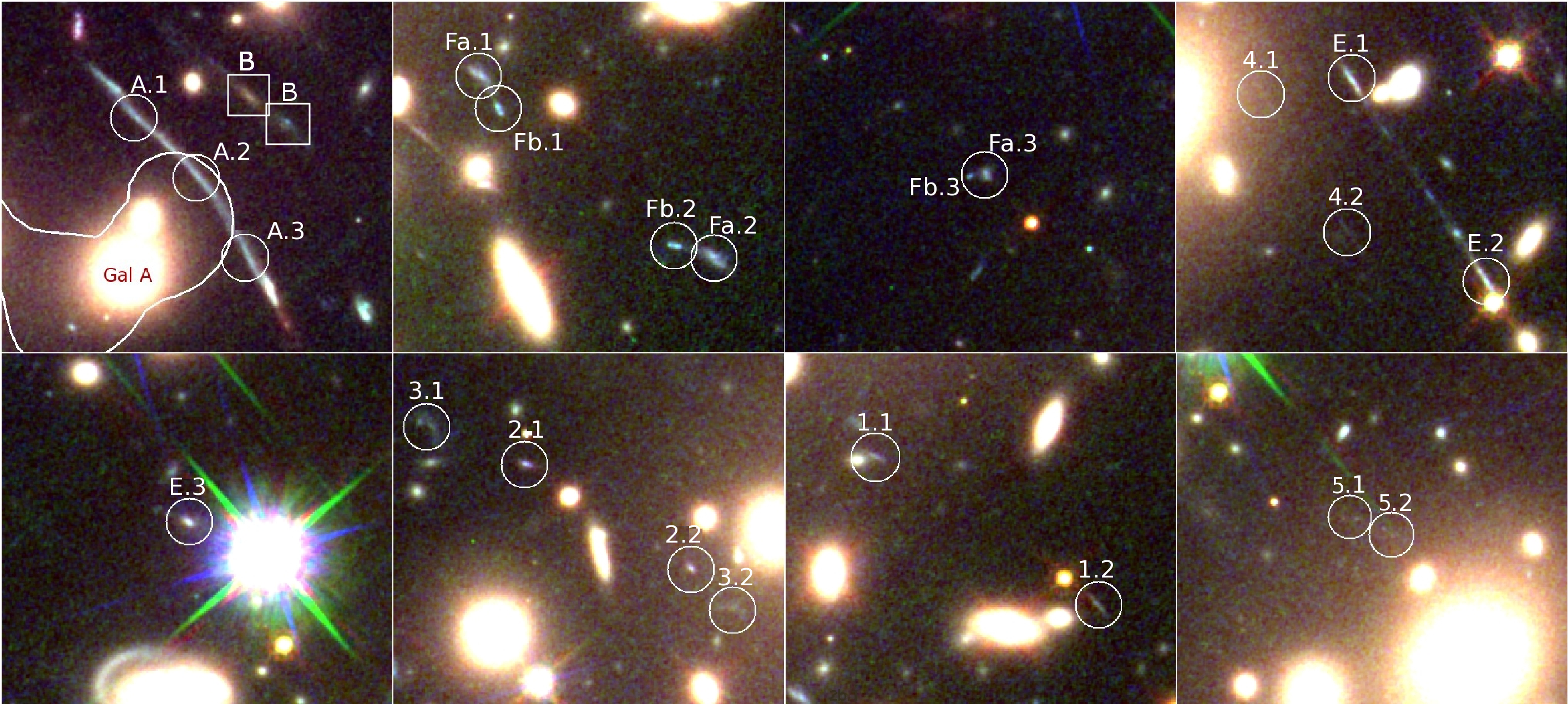}
\caption{The multiple images identified in the main cluster area with the HST images, as shown in  Figure~\ref{MainClump}. System K is not shown since it is invisible in ACS/HST. The size of each box is $12\arcsec\times12\arcsec$, North is up and East is left.}
\label{MainClumpMulti}
\end{figure*}

   \begin{figure*}
   \centering
\includegraphics[width=0.95\textwidth]{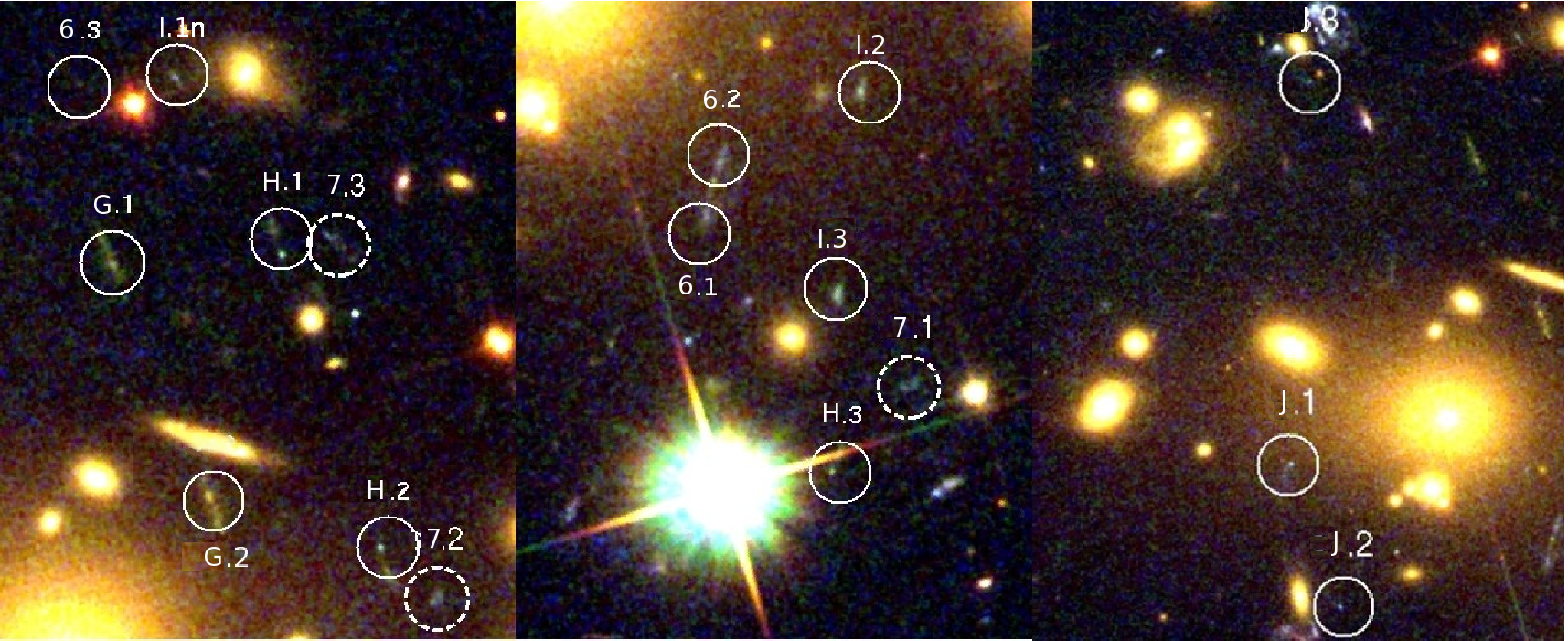}
\caption{The multiple images identified in the sub cluster area with the ACS images, as shown in Figure~\ref{SubClump}.  Multiple images  marked with dashed line circles mark the predicted but not confirmed positions of counter images. System 7 with  dashed circle, is a multiply-imaged candidate and is not a part of model constraints due to large color uncertainties. The size of each box is $20\arcsec\times12\arcsec$, North is up and East is left.}
\label{SubClumpMulti}
\end{figure*}

   \begin{figure*}
      \centering

\includegraphics[width=0.95\textwidth]{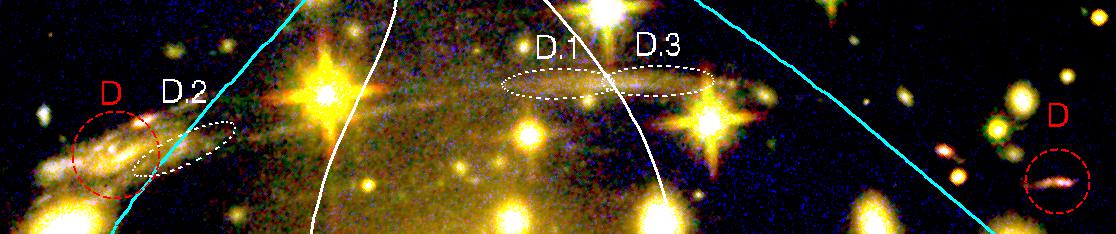}

\caption{The multiply-imaged system D, a giant arc, identified in the  cluster area with the ACS images, as shown in Figure~\ref{MainClump}.  Multiple images  marked with red dashed line circles mark the positions of  images of system D as reported by \citet{Bradac:2006}. White ellipses mark the positions of multiple images of the D system predicted by our strong lensing mass model. The white line is the critical line at the model redshift  corresponding to this candidate system ($z=3.23$). Cyan lines mark limits of multiple image occurrence as predicted by our model.}
\label{DMulti}
\end{figure*}

 We have reviewed each strongly lensed candidate proposed by \cite{Mehlert:2001, Bradac:2006, Bradac:2009} by checking 1) the morphology  2) color agreement and 3) the consistency with the lensing mass model prediction. In the end, we have only used those systems that have passed all three tests.  By color agreement, we call images that ``color distance''  is less than 5.5.  ``Color distance''   is defined as 
 \begin{equation}
 [\sum_{ij}\mid c_i-c_j\mid/\sqrt{cerr_i^2+cerr_j^2}]/N,  
 \end{equation}
where $c_i,c_j$ are colors and $cerr_i, cerr_j$ are color errors.

In order to avoid confusion between our and previous multiple image identification, we choose to keep alphabetical notation (A-L), for only those multiple image systems  that were already reported by  \citet{Bradac:2006, Bradac:2009}, while for our newly identified images we use numerical notation (1-7).
Altogether, our set of constraints is quite different from the one of \cite{Bradac:2009}. 
Indeed, we have rejected 3 multiple-image systems (B, C and L)  and changed system A, D and E, for the following reasons:
 \begin{itemize}
 \item %----------
 system A: \cite{Bradac:2006} matched two symmetric images A1 and A2 forming a giant arc, whereas the expected third image A3 is not detected. Moreover, the analysis of Spitzer/MIPS and Herschel data of the Bullet cluster performed by \cite{Rex:2010} shows that only A1 is detected in the Far-IR (source HLS12 from this paper) and therefore the identification of the system is incorrect (see Table~\ref{Asystem}). To reconciliate the identification with the far-infrared observations, we interpret the giant arc as the merging of 3 images crossing the critical line at z=3.24 (system A.1, A.2 and A.3 in our notation) (see Figure~\ref{MainClumpMulti} and Table~\ref{mainsystem});
 
%----------
\item  system B: according to our model (which now takes into account triply-imaged system A) B is a single imaged arc, also the colors between B1 and B2 used by  \citet{Bradac:2006}  are  clearly different as seen in Figure~\ref{MainClumpMulti} and Table~\ref{BCLsystem};
\item system C: the predicted third image is not detected  (see also Table~\ref{BCLsystem});
%, and the colors of the two images are in disagreement  \citep[the Northern C image is detected with Spitzer/MIPS, however the Southern is not, ][]{Perez:2010};

 \item system D: is problematic due to extended morphology and the large uncertainty in locating the different multiple image centers, thus we do not add it to the set of the systems that constrain the mass model, instead we use the mass model to predict two new image positions D.1 and D.3 and  their redshift (see Table~\ref{mainsystem}, Figure~\ref{MainClump} and \ref{DMulti}); 
  \item  system E:   (in our notation system 3), we find  its third counter image E.3;
\item  system L:   the two images L1 and L2 reported by \citet{Bradac:2009} have slightly color inconsistency (L1 is brighter than L2 by $\sim$0.5 mag in all the filters except for F606W filter where L2 is detected with $>5$ $\sigma$ and L1 is not  detected with $<1.5$ $\sigma$),  ``color distance''  is 3.6, additionally,  positions of multiple images is  entirely excluded by the geometry of our model  (RMS $>10\arcsec$).

  \end{itemize}

In the sub-cluster, we have also excluded/added the following images of the I, J and G systems:
\begin{itemize}
\item    old image  I1, identified by \citet{Bradac:2009}  has a significantly different color  than images I2 and I3, instead we have found an object 'new I1n' that better fits the position and color of the system (see Table~\ref{Isystem} and Figure~\ref{SubClump});
\item  according to the geometry of our mass model, the old G3 identification can not belong to the same system as images G1 and G2. The model  predicts  position of the third  counterpart G3 of the system directly on the bright star, south of the cluster.  We displayed the new G3n position predicted by the model in Figure~\ref{SubClump}; 

\item  we believe that the old G3 is a "straight" arc made of 2 merging images (named '6.1' and '6.2' in our notation, see Figure~\ref{SubClump}) with an identified third counterpart in the North part of the sub-cluster (6.3);

\item  we identified the third image of the system J (J.3 as shown in Figure~\ref{SubClump}).

\end{itemize}

\begin{table}
\centering

\begin{tabular}{lcccc}
\hline\hline
Name & F606W & F435W-F606W & F606W-F814W\\
\hline
I.1$^a$& $26.49\pm0.11$&$ 0.18\pm0.18$& $-0.22\pm0.15$\\
I.1n$^b$& $26.36\pm 0.14$&$0.47\pm0.11$&$0.17\pm0.07$\\
I.2& $25.55\pm 0.07$&$0.50\pm0.07$&$0.08\pm0.04$\\
I.3& $25.60\pm0.05$&$0.49\pm0.06$&$0.03\pm0.04$\\
\hline
\end{tabular}
\caption{Photometry of the components of the multiply-imaged system I. We note the strong color difference
in F606W-F814W colors for the old I.1$^a$ by \citet{Bradac:2006} and new, this paper I.1n$^b$ identification. See also Figure~\ref{SubClump} and \ref{SubClumpMulti}}.
\label{Isystem}
\end{table}

The  three remaining multiply-imaged systems from  \cite{Bradac:2009} F  and K in the main cluster and H in the sub-cluster,  were  included in our set of constraints without changes.

Finally, thanks to the new deep WFC3 and ACS images, we have identified 6 new secure multiply-imaged systems, five new systems (1-5) in the main cluster and one new system (6) in the sub-cluster.  
We have also identified system 7 in the sub-cluster, however due to large uncertainties in the color of this system, we do not use it as a part of our model constraints, instead we present it only as a possible multiply-imaged candidate.

In Figure~\ref{MainClumpMulti}, \ref{SubClumpMulti} and \ref{DMulti} we show postage stamps of
all multiple images (except for system K that is not visible in HST/ACS and barely detected in HST/WFC3, see \citet{Bradac:2009} for a Spitzer postage stamps), their exact locations, photometry, magnitudes, colors, magnifications and redshifts are given in Table~\ref{mainsystem}.

In total, we use 14 strongly lensed  systems  (9 systems in the main cluster region and 5 systems in the sub-cluster region), three of those systems (A, K and H) have measured spectroscopic redshifts.
System A is the bright giant arc with previously measured spectroscopic redshift at z=3.24 by \cite{Mehlert:2001}, (see Figure~\ref{MainClumpMulti}), system K  is an IRAC bright submm source, dusty galaxy, with well measured redshift z=2.79 by \citet{Gonzalez:2010}, and system  H at redshift z=2.99 was measured in this paper with FORS/VLT (see Section 2.3).

\begin{table*}
\begin{center}
\begin{tabular}{lcccccccc}
\hline\hline
  &R.A.&Dec&F606W&MIPS&F606W-F775W&F775W-F850LP&F850LP-F110W&F110W-F160W\\
\hline
A&104.63332& -55.941377&$25.71\pm0.04$&0.06 mJy&$0.39 \pm 0.03 $&$ 0.10 \pm 0.02 $&$ 0.11 \pm 0.02 $&$ 0.22 \pm 0.01 $\\
A&104.63020& -55.943620&$25.00\pm0.04$&$<$0.016 mJy&$0.86 \pm 0.03 $&$ 0.04 \pm 0.01 $&$ 0.01 \pm 0.01 $&$ 0.33 \pm 0.01 $\\
\hline
\end{tabular}
\caption{Photometry of the components of the multiply-imaged system A as identified by \citet{Bradac:2009}. We note the strong color difference between MIPS and other filters  OF the old A system identification by \citet{Bradac:2006} and good color agreement of new system A identification (see Table~\ref{mainsystem}). See also Figure~\ref{MainClump} and \ref{MainClumpMulti}  }
\label{Asystem}
\end{center}
\end{table*} 
\begin{table*}
\begin{center}
\begin{tabular}{lcccccc}
\hline\hline
  &R.A.&Dec&F606W-F775W&F775W-F850LP&F850LP-F110W&F110W-F160W\\
\hline\hline
B&104.62968&-55.9418082&$0.58 \pm 0.09 $&$ 0.05\pm 0.07 $&$ -0.01 \pm 0.05$&$ 0.00 \pm 0.03 $\\
B&104.63047&-55.9414692&$0.90 \pm 0.12 $&$ 0.30 \pm 0.07 $&$  0.36 \pm 0.04$&$ 0.20 \pm 0.01 $\\
C  & 104.63729 &-55.942493&$0.67 \pm 0.23 $&$ 0.32 \pm 0.16 $&$ 0.56 \pm 0.11 $&$ 0.36 \pm 0.06 $\\
C  & 104.63339 &-55.945603&$0.23 \pm 0.05 $&$ 0.10 \pm 0.06 $&$ 0.49 \pm 0.04 $&$ 0.43 \pm 0.02 $\\
L& 104.64339& -55.963841&$< 2.93 $&$ 0.43 \pm 0.12 $&$ 0.69 \pm 0.08 $&$ 0.28 \pm 0.04 $\\
L& 104.650356 &-55.961562&$1.35 \pm 0.59 $&$ 0.79 \pm 0.21 $&$ 0.77 \pm 0.12 $&$ 0.27 \pm 0.05 $\\

 % D.3          &104.64623&  -55.944115&  $-0.10\pm0.15$& $0.06\pm0.10$&  $0.45\pm0.07$&$  0.29\pm0.04$\\
%  D.1          &104.63943&  -55.947552& $-0.49\pm0.35$&$0.49\pm0.33$&$0.13\pm0.22$&$0.30\pm0.17$\\
%  D.2             &104.63981&  -55.947308&  $ 0.08\pm0.21$&$ 0.19\pm0.10$& $0.17\pm0.07$&$ 0.17\pm0.04$\\

  %D1.paper     &104.63943&  -55.947552&$-0.49\pm0.35$&$0.49\pm0.33$&$0.13\pm0.22$&$0.30\pm0.17$\\
 % D2.paper     &104.63823&  -55.948384&$-0.13\pm0.10$&$0.99\pm0.14$&$-0.03\pm0.52$&$0.37\pm0.11$\\
  %D3.paper     &104.64794&  -55.943838&$0.45\pm0.25$&$0.04\pm0.22$&$0.30\pm0.12$&$0.20\pm0.31 $\\ 
 % Dbradac  $^a$&  104.63546 &-55.951920 &$ -0.15 \pm0.13$& $-0.11\pm0.09$& $0.43\pm0.06$ &$ -0.88\pm0.03$\\
  %Dbradac  $^a$&  104.64713 &-55.943712 &$ 0.31\pm0.11$&$ -0.12\pm0.06$& $0.77\pm0.04$&$  -0.42\pm0.02$\\
  %D.2          &104.63981&  -55.947308&  $ 0.1995 \pm 0.0807 $&$ 0.0232 \pm 0.0583 $&$ 0.4104 \pm 0.0367 $&$0.2655 \pm 0.0107 $\\
  %D.1          &104.63943&  -55.947552& $ -0.3129 \pm 0.1442 $&$ -0.1369 \pm 0.1555 $&$ 0.8004 \pm 0.1009 $&$0.3586 \pm 0.0199 $\\
  %D.3          &104.64623&  -55.944115&  $ -0.0284 \pm 0.0652 $&$ 0.17 \pm 0.0559 $&$ 0.5678 \pm 0.033 $&$0.3159 \pm 0.0099 $\\

  D &  104.64713 &-55.943712 &$  -0.25 \pm 0.10 $&$ 0.63 \pm 0.08 $&$ 0.32 \pm 0.04 $&$0.18 \pm 0.02 $\\
  D &  104.63546 &-55.951920 &$ -0.75 \pm 0.17 $&$ 0.17 \pm 0.22 $&$ 0.39 \pm 0.13 $&$0.96 \pm 0.02 $\\
\hline
\end{tabular}
\caption{ Photometry of the rejected systems B,C, D and L as identified by \citet{Bradac:2009}.  }
\label{BCLsystem}
\end{center}
\end{table*} 
%==============================================================================================================================
\section{Lensing Methodology}
%==============================================================================================================================
\subsection{Overview}
\begin{table*}
\begin{center}
\hspace*{-1.2cm}
\resizebox{20cm}{!} {
\begin{tabular}{lcccccccccccc}
\hline\hline
System  &R.A.&Dec&F606W&F606W-F775W&F775W-F850LP&F850LP-F110W&F110W-F160W&$\mu$&$z_{\rm
phot}$&$z_{\rm m}^a$&$z_{\rm spec}$\\
&(deg)&(deg)& \\
\hline
\multicolumn{3}{l}{Main Clump}\\
\hline
A.1                          &104.63293     &-55.941725 &$25.59\pm0.07$&$0.64\pm0.09$&$0.02\pm0.09$&$0.13\pm0.08$&$0.41\pm0.05$&$29.56\pm8.97$&$3.98_{-0.07}^{+0.04}$ &--&3.24 \\
A.2                          &104.63158     &-55.942454 &$25.65\pm0.05$&$0.51\pm0.09$&$0.09\pm0.09$&$0.21\pm0.07$&$0.42\pm0.04$&$26.22\pm6.68$&$3.09_{-0.62}^{+0.45}$ &--&-- \\
A.3                          &104.63055     &-55.943405 &$25.67\pm0.06$&$0.66\pm0.11$&$0.02\pm0.09$&$0.02\pm 0.08$&$0.37\pm0.06$&$12.48\pm0.90$&$3.17_{-0.14}^{+0.03}$& -- & -- \\
Fa.1$^b$               &104.65210     &-55.956245&$24.63\pm0.06$&$0.30\pm0.08$&$-0.07\pm0.09$&$0.15\pm0.08$&$0.42\pm0.05$&$9.77\pm0.44$&$2.86_{-0.25}^{+0.35}$& $2.14\pm0.19$&-- \\
Fa.2                        &104.64704     &-55.958497&$24.12\pm0.05$&$0.22\pm0.03$&$-0.08\pm0.03$&$0.18\pm0.03$&$0.41\pm0.05$&$11.11\pm0.72$&$2.70_{-0.24}^{+0.61}$&--&--\\
Fa.3                        &104.66515     &-55.951289&$25.33\pm0.11$&$0.37\pm0.15$&$-0.01\pm0.15$&$0.08\pm0.14$&$0.52\pm0.09$&$4.89\pm0.22$&$2.96_{-0.27}^{+0.20}$&--&--  \\
Fb.1$^b$                &104.65165     &-55.956641&$25.31\pm0.09$&$0.30\pm0.08$&$-0.07\pm0.09$&$0.14\pm0.08$&$0.39\pm0.08$&$14.91\pm1.11$&$3.10_{-0.93}^{+0.54}$&--&--\\
Fb.2                         &104.64786     &-55.958292&$25.09\pm0.12$&$0.24\pm0.04$&$-0.03\pm0.04$&$0.06\pm0.04$&$0.13\pm0.10$&$15.02\pm0.96$&$2.85_{-0.19}^{+0.35}$&--&--\\
Fb.3                         &104.66544     &-55.951301&$26.61\pm0.36$&$0.45\pm0.11$&$-0.08\pm0.11$&$0.04\pm0.10$&$0.27\pm0.09$&$4.77\pm0.21$&$3.98_{-1.62}^{+0.40}$& --&--  \\
E.1                          &104.64242     &-55.948720&$25.52\pm0.05$&$0.49\pm0.08$&$0.11\pm0.07$&$0.15\pm0.06$&$0.43\pm0.08$&$15.96\pm1.36$&$3.50_{-0.45}^{+0.38}$&$3.17\pm0.27$&--  \\
E.2                          &104.63954     &-55.951165&$25.51\pm0.04$&$0.55\pm0.08$&$0.22\pm0.06$&$0.11\pm0.05$&$0.37\pm0.05$&$16.72\pm1.30$&$3.16_{-0.50}^{+0.39}$&--&--   \\
E.3                           &104.65434     &-55.944392
&$25.77\pm0.04$&$0.45\pm0.07$&$0.33\pm0.06$&$0.20\pm0.05$&$0.42\pm0.06$&$  7.13\pm0.40$&$2.96_{-0.02}^{+0.01}$&--&-- \\
K.1                          &104.65864     &-55.950557   &--&--    &--&--&--&--&--                        &--&  2.79 \\
K.2                          &104.65459     &-55.951876   & --&--    &--&--&--&--&--                         &--&  --\\
K.3                          &104.63929     &-55.958032   & --&--    &--&--&--&--&--                         &--&  --\\
1.1                           &104.65049     &-55.953339&$27.20\pm0.13$&$0.25\pm0.17$&$0.19\pm0.25$&$0.89\pm0.18$&$0.17\pm0.04$&$15.65\pm1.20$&$2.03_{-0.54}^{+0.96}$&$1.21\pm0.23$&--\\
1.2                           &104.64572     &-55.955109&$26.83\pm0.11$ &$0.29\pm0.09$&$0.30\pm0.07$&$0.50\pm0.05$&$0.24\pm0.05$&$11.20\pm0.57$&$1.39_{-0.39}^{+1.07}$&--&-\\
1.3*                           &104.65706     &-55.950866   &--&-- &--&--&--&--&                      --&-- \\
2.1                           &104.65560     &-55.948760&$25.62\pm0.06$&$0.32\pm0.21$&$0.27\pm0.26$&$0.21\pm0.21$&$0.51\pm0.11$&$10.61\pm0.40$&$2.96_{-0.02}^{+0.02}$&$2.91\pm0.25$&--   \\
2.2                           &104.65203     &-55.950014&$24.59\pm0.06$&$0.56\pm0.12$&$0.24\pm0.09$&$0.21\pm0.08$&$0.49\pm0.05$&$14.18\pm0.94$&$3.09_{-0.29}^{+0.73}$&--&-  \\
2.3                           &104.63801     &-55.956311&$26.35\pm0.07$&$0.45\pm0.16$&$0.44\pm0.13$&$0.25\pm0.10$&$0.57\pm0.08$&$4.79\pm0.21$&$2.74_{-0.02}^{+0.10}$&--&- \\
3.1                           &104.65770     &-55.948271&$24.74\pm0.05$&$0.61\pm0.18$&$-0.01\pm0.07$&$-0.09\pm0.05$&$0.23\pm0.02$&$9.37\pm0.54$&$4.16_{-0.25}^{+0.22}$&$4.06\pm1.23$&--\\
3.2                           &104.65118     &-55.950472&$26.74\pm0.08$&$0.95\pm0.32$&$-0.22\pm0.17$&$-0.26\pm0.15$&$0.38\pm0.12$&$4.73\pm0.14$&$4.32_{-0.14}^{+0.28}$&--&--\\
3.3                           &104.63565     &-55.957061   &$>28.50$&$>1.51$     &$-0.34\pm0.07$&$-0.05\pm0.27$&$0.53\pm0.24$&$4.94\pm0.22$ &$4.53_{-0.55}^{+0.42}$&-&-\\
4.1                           &104.64423     &-55.949050&$27.58\pm0.38$&$0.21\pm0.12$&$0.57\pm0.04$&$-0.29\pm0.04$&$-0.08\pm0.05$&$22.96\pm2.11$&$2.03_{-0.42}^{+1.1}$&$2.00\pm0.12$&--\\
4.2                           &104.64251     &-55.950576&$26.97\pm0.28$&$0.28\pm0.14$&$0.36\pm0.07$&$-0.33\pm0.05$&$-0.20\pm0.04$&$20.04\pm1.71$&$0.17_{-0.12}^{+2.28}$&--&--\\
4.3*                           &104.65166     &-55.944823   &--&--&--&--&--&--&--                        &--&-- \\
5.1                           &104.64961     &-55.947382&$27.44\pm0.23$&$1.14\pm0.21$&$-0.72\pm0.19$&$-0.06\pm0.17$&$0.26\pm0.35$&$4.70\pm0.20$&$2.10_{-0.21}^{+0.38}$&$2.54\pm0.34$&--\\
5.2                           &104.64928     &-55.947412&$27.56\pm0.45$&$0.90\pm0.18$&$-0.31\pm0.12$&$-0.40\pm0.10$&$0.21\pm0.26$&$4.75\pm0.20$&$2.00_{-0.18}^{+0.53}$&--&--\\
5.3*                           &104.63731     &-55.953223     &--
 &--    &--&--&--&--                        & --& -- \\
%D.1$^c$     &104.63940     &-55.947582&$26.79\pm0.24$&$-0.49\pm0.35$&$0.49\pm0.33$&$0.13\pm0.22$&$0.30\pm0.17$&$50.19\pm5.42$&$2.56_{-0.38}^{+0.74}$&$3.23\pm0.42$&--\\
%D.2            &104.63843     &-55.948325&$26.74\pm0.20$&$-0.13\pm0.10$&$0.99\pm0.14$&$-0.03\pm0.52$&$0.37\pm0.11$&$49.83\pm10.34$&$2.34_{-0.45}^{+0.24}$&--&--\\
%D.3            &104.64806     &-55.943810&$26.59\pm0.22$&$0.45\pm0.25$&$0.04\pm0.22$&$0.30\pm0.12$&$0.20\pm0.31 $&$9.96\pm3.23$&$2.13_{-0.89}^{+0.85}$&--&--\\
  D.2 $^c$         &104.63981&  -55.947308&$26.56\pm0.20$& $ -0.19 \pm 0.08 $&$ 0.02 \pm 0.06 $&$ 0.41 \pm 0.04 $&$0.27 \pm 0.01$ &$49.83\pm10.34$&$2.34_{-0.45}^{+0.24}$&--&--\\
  D.1          &104.63943&  -55.947552& $26.79\pm0.24$&$ -0.31 \pm 0.14 $&$ -0.13 \pm 0.16 $&$ 0.80 \pm 0.10 $&$0.36 \pm 0.02 $&$50.19\pm5.42$&$2.56_{-0.38}^{+0.74}$&$3.23\pm0.42$&--\\
  D.3          &104.64623&  -55.944115&$25.52\pm0.22$&  $ -0.03 \pm 0.07 $&$ 0.17 \pm 0.06 $&$ 0.57 \pm 0.03 $&$0.32 \pm 0.01 $&$49.96\pm3.23$&$2.13_{-0.89}^{+0.85}$&--&--\\

\hline
\end{tabular}}
\caption{The multiply-imaged systems used to constrain the model, with their ID, centroid position, brightness,  colors, linear magnification and predicted or spectroscopically measured redshifts.
The systems and their properties, along with references to the papers reporting their redshifts, are given in Section 3. 
Images marked * are not detected, thus their positions are just predictions from lens model.  We do not report photometry on system K since it is smm galaxy, invisible in ACS/HST.
$^a$ Redshift estimation inferred from the mass model when all spectroscopically confirmed multiply-imaged systems have been included in the optimization. 
$^b$  System Fa and Fb are most probably gravitationally bounded.
$^c$  System  D is not included in the model optimization, it is a multiply-imaged candidate.
Numbers (.1, .2 ,etc) denote the different images of each set of multiple images. Each multiple image is  presented in Figure~\ref{MainClumpMulti}, except system K that is invisible in optical and system D  that has not been included in the model optimization  is  presented in Figure~\ref{DMulti}.  }
\label{mainsystem}
\end{center}
\end{table*} 

\begin{table*}
\begin{center}
\hspace*{-1.4cm}
\resizebox{20cm}{!} {
\begin{tabular}{lcccccccccc}
\hline\hline
System  &RA.&Dec&F606W&F435W-F606W&F606W-F814W&F435W-F814W&$\mu$&$z_{\rm
phot}$&$z_{\rm m}^a$&$z_{\rm spec}$\\
&(deg)&(deg)& \\
\hline
\multicolumn{3}{l}{Sub Clump}\\
\hline
H.1           &104.56305   &-55.939755&$26.59\pm0.08$&$0.65\pm0.11$&$-0.21\pm0.07$&$0.43\pm0.12$&$6.48\pm0.67$&$3.3_{-1.2}^{+0.2}$&--&2.99 \\
H.2           &104.56145   &-55.942423 &$26.78 \pm0.08$&$0.78\pm0.14$&$-0.40\pm0.09$&$0.36\pm0.16$&$9.62\pm0.87$&$3.3_{-0.7}^{+0.2}$&--&-- \\
H.3           &104.56202   &-55.947717  &$26.57\pm0.12$&$1.00\pm0.15$&$-0.29\pm0.06$& $0.29\pm0.14$&$11.74\pm1.53$&$2.7_{-1.3}^{+0.9}$&-- &-- \\
I.1n            &104.56478   &-55.938146  &$26.36\pm 0.14$&$0.47\pm0.11$&$0.17\pm0.07$&$0.64\pm0.11$&$5.31\pm0.50$ &$2.4_{-1.4}^{+0.9}$&--&--  \\
I.2             &104.56155   &-55.944252  &$25.55\pm 0.07$&$0.50\pm0.07$&$0.08\pm0.04$&$ 0.59\pm0.07$&$20.25\pm3.38$ &$2.6_{-1.5}^{+0.7}$&$3.24\pm0.13$&-- \\
I.3             &104.56192   &-55.946101 &$25.60\pm0.05$&$0.49\pm0.06$&$0.03\pm0.04$& $0.52\pm0.06$&$15.91\pm1.53$&$2.6_{-1.5}^{+0.7}$&--&--  \\
J.1            &104.57038   &-55.944036  &$26.81\pm0.13$&$-0.11\pm0.08$&$-0.47\pm0.09$&$-0.58\pm0.10$&$24.49\pm4.01$&$1.6_{-0.6}^{+1.2}$&$2.02\pm0.25$&--  \\
J.2            &104.56917   &-55.946003  &$27.03\pm0.12$&$-0.32\pm0.10$&$-0.19\pm0.10$&$-0.50\pm0.11$&$19.33\pm1.99$&$1.0_{-0.0}^{+1.8}$&--&--\\
J.3            &104.56993   &-55.938772  &$27.99\pm0.55$&$-0.17\pm0.21$&$-0.37\pm0.16$&$-0.20\pm0.18$&$5.39\pm0.48$&$0.9_{-0.9}^{+2.3}$&--&--\\
G.1          &104.56580   &-55.939857  &$25.31\pm0.05$&$>1.85$&$0.40\pm0.05$&$>2.25$&$11.60\pm1.56$&$3.3_{-2.3}^{+1.5}$&$0.73\pm0.09$&--\\
G.2          &104.56424  &-55.941963  &$25.27\pm0.06$&$>1.73$&$0.57\pm0.06$&$>2.16$&$13.46\pm1.04$&$2.5_{-0.8}^{+1.9}$&--&--\\
G.3n*          &104.56471   &-55.947724  &--&--&--&--&--&-- \\
6.1          &104.56406   &-55.945386&$26.05\pm0.08$&$-0.14\pm0.09$& $0.25\pm0.07 $&$0.11\pm0.08$&$40.90\pm4.88$&$0.9_{-0.6}^{+1.1}$&$2.51\pm0.12$&--\\
6.2          &104.56384   &-55.944904       &$26.17\pm0.11$&$-0.08\pm0.09$& $0.15\pm0.08$&$0.07\pm0.09$&$31.30\pm5.68$&$0.8_{-0.8}^{+1.9}$&--&--\\
6.3*          &104.56627   &-55.938208&$>28.45$&--&--&--&$4.86\pm0.43$&--&--\\
$7.1^b $    &104.56084  &-55.946952  & $ 26.89\pm0.15$&$0.07\pm0.35$&$-0.36\pm0.45$&$-0.29\pm0.23$&$22.71\pm3.39$&$1.4_{-0.8}^{+1.9}$&$3.00\pm0.19$&--\\
7.2          &104.56054   &-55.942898     &$26.54\pm0.12$&$0.15\pm0.22$& $0.03\pm0.32$&$0.18\pm0.13$&$16.79\pm1.90$&$2.2_{-1.5}^{+1.5}$&--&--\\
7.3          &104.56230   &-55.939588     &$26.89\pm0.13$&$0.03\pm0.34$& $0.25\pm0.29$&$0.28\pm0.38$&$10.94\pm0.78$&$0.5_{-0.4}^{+3.0}$&--&--\\
\hline
\end{tabular}}

\caption{The multiply-imaged systems used to constrain the model, with their centroid position, brightness,  colors, linear magnification and predicted or spectroscopically measured redshifts.
The systems and their properties, along with references to the papers reporting their redshifts, are given in Section 3. Images marked * are not detected, thus their positions are just  predictions from lens model. 
$^a$ Redshift estimation inferred from the mass model when all spectroscopically confirmed multiply-imaged systems have been included in the optimization. 
$^b$  System 7  is not included in the model optimization, it is multiply-imaged candidate.
Numbers (.1, .2 ,etc) denote the different images of each set of multiple images, each presented in Figure~\ref{SubClumpMulti}, except system 7 that has not been included in the model optimization.  }
\label{subsystem}
\end{center}
\end{table*}

The strong-lensing mass reconstruction  is based on the Bayesian Monte Carlo Markov chain (MCMC) method implemented in the  \textsc{LENSTOOL} \footnote{See http://projets.oamp.fr/projects/lenstool/wiki } software  \citep{Kneib:1996, Jullo:2007, Jullo:2009}. 

The  mass distribution of the Bullet cluster is considered here as a superposition of three cluster-scale dark matter clumps (two in the main cluster and one in the sub-cluster),  the BCGs, the intracluster gas and the individual galaxies. 

The light distribution of the  main  cluster indicates cluster-scale dark matter bimodality.  Nonetheless, we have checked the alternative possibility of  main  cluster consisting of only one dark matter clump. This alternative can not reproduce the position of the multiple images with as high precision as the two-clump model. It gives a significantly worse fit to the data (RMS = 1.5$\arcsec$  vs RMS = 0.2\arcsec), confirming the existence for two large scale dark matter halos in the main clump.
Therefore, the dark matter clumps are called DM1 and DM2 in the bimodal main clump and DM3 in the sub-clump.

All  dark matter clumps and galaxies were parameterized as dual Pseudo Isothermal Elliptical mass distributions  dPIE \citep{Limousin:2005}.
The  dPIE is described by seven parameters:  redshift, central position ($x_{c},y_{c}$), ellipticity 
$\epsilon=\frac{a^2-b^2}{a^2+b^2},$
(with a and b being semi-major and semi-minor axis, respectively),
the position angle $\theta$, a core radius $r_{\rm core}$, a truncation radius $r_{\rm cut}$  and a fiducial velocity dispersion $\sigma$. 
The two scale radii, $r_{\rm core}$ and $r_{\rm cut}$,  define changes in the slope of the  dPIE density profile:
\begin{equation}
\rho(r)=\frac{\rho_0}{(1+r^2/r^2_{\rm core})(1+r^2/r^2_{\rm cut})},
\end{equation}
where $\rho_0$ is a central density.
The profile is flat in the inner region, then isothermal ($\rho\sim r^{-2}$) between $r_{\rm core}$ and $r_{\rm cut}$, and steeply decreasing ($\rho \sim r^{-4}$) beyond $r_{\rm cut}$.
Thanks to its extra degree of freedom compared to a NFW potential, the dPIE potential is more flexible in modeling complex galaxy clusters, such as 1E 0657-56. Moreover,  the  dPIE profile is specially suitable to model  galaxies. Indeed, several studies, based on dynamics of stars, globular clusters and X-ray halos have shown that early-type galaxies are isothermal in their inner  parts \citep{Koopmans:2006, Oguri:2007, Gerhard:2001, Peng:2004}, with no significant evolution with redshift up to $z\sim1$,  this is also true for cluster members \citep[see][]{Natarajan:1998,Natarajan:2009}.
The  dPIE has been successfully used to model galaxy clusters \citep[e.g.,][]{Kneib:1996, Richard:2007}, as well as early-type galaxies \citep[e.g.,][]{Natarajan:1998, Limousin:2007a}

In the optimization procedure the  dPIE parameters of cluster scale DM halos (the central position, ellipticity, velocity dispersion, core radius  and the position
angle) were allowed to  vary freely. 
In case of the cluster galaxies, the position, ellipticity, and orientation were matched to that of the
light distribution as measured by \textsc{SExtractor}. The velocity dispersions, core and cut-off radii of cluster members were scaled with their luminosity using common scaling relations (see Section 4.2).

\begin{figure}
   \centering
\includegraphics[scale=0.55]{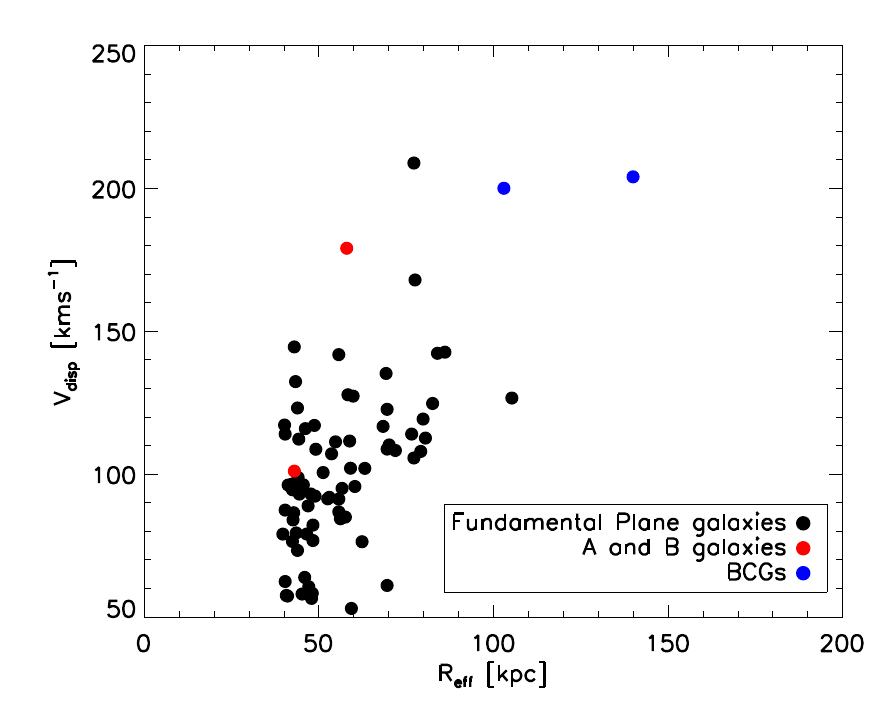}
\caption{ The galaxy members identified using color-magnitude cut and comparison of the properties with separately modeled galaxies A and B (red) and BCGs (blue).
}
\label{GalaxyMembers}
\end{figure}

BCGs and the 2 cluster galaxies (see Figure~\ref{GalaxyMembers})  that are in the vicinity of multiple images were fitted individually, the same approach was used in lens modeling by \citet{Suyu:2010,Limousin:2008, Limousin:2007b, Richard:2010, Richard:2010a}. 
Firstly, because both BCG and galaxies  that are in the vicinity of multiple images have strong influence on the multiple images position.
Secondly, because BCGs are likely distinct galaxy population from cluster ellipticals, hence they do not follow common scaling laws
 \citep{Natarajan:1998}. The two galaxies that are used explicitly in the optimization procedure are: galaxy A (104.63308, -55.943594) which strongly affects  the brightest tangential arc (system A) and galaxy B (104.65626, -55.950795) which affects the multiply-imaged systems 1 and K  (see  Figure~\ref{MainClump} and \ref{MainClumpMulti}).

The baryonic matter content of galaxy clusters is dominated by the X-ray emitting intra-cluster gas, the mass of which reaches 10-15\% \citep{David:1993, Neumann:2001, Vikhlinin:2006, LaRoque:2006} of the total mass. As the emissivity of the X-ray emitting gas is proportional to the square of its density, the gas mass profile in a cluster can be precisely determined from X-ray data. In the case of the Bullet cluster there is a significant offset between the gas and dark matter distribution. Because of this offset,  including the gas mass as a separate component is important for accurate modeling of the total mass distribution.
Therefore, we have included the intra-cluster gas (without optimization of this component) in our total mass model  of the Bullet cluster (see Section 4.3) using the X-rays measurements performed by  \cite{Ota:2004, Markevitch:2002}.

Using the observational constraints (namely multiply image positions and photometric redshifts given in Table~\ref{mainsystem} and~\ref{subsystem})
we have optimized the parameters of the mass components: the DM clumps, 
the BCGs and the individual galaxies (see Table~\ref{modelresults}).
As a starting point, we have used a set of initial parameters (centroid, ellipticity and position angle) based on the visible component \citep{Limousin:2008},  that were then iteratively optimized.

For each image, 
 marginalising over the photometric redshift (if no spectroscopic redshift was known)
we find its RMS (root-mean-square) value for its position in the  image plane, given by
$\rm RMS =\sqrt{\frac{1}{n}\sum^{n}_{j=1} (X^{\rm j}_{\rm obs} - X^{\rm j}_{model})^2}$,
where n is the number of images for the system, X$_{\rm model}$ is the predicted by model position in the
image plane and X$_{\rm obs}$ the observed position in the image plane (See Table \ref{tab-rms}). The overall RMS is defined by summing and averaging over all the images for all the systems. A detailed overview of the \textsc{LENSTOOL} software and discussion of parameters uncertainty  can be found in \cite{Jullo:2007}. 
Thanks to the  parallelized version of \textsc{LENSTOOL} the optimization could be efficiently performed  in the image plane  similarly as done in \cite{Limousin:2011}.

%===================================================================================================================
\subsection{Scaling relation of elliptical cluster members}
%==============================================================================================================================

We lack sufficient sensitivity to constrain the detailed mass profile for individual cluster galaxies. Thus, in general,  cluster modeling uses Faber-Jakson scaling relation (FJR) \citep{Faber:1976} to scale the galaxy members. It assumes that  all galaxies in the cluster have the same M/L ratio  \citep{Natarajan:1997,Natarajan:1998, Limousin:2007b, Oguri:2010,Limousin:2005, Kneib:2003, Richard:2009, Jullo:2010}.
 FJR is however, an  empirical relation with a quite large scatter. 
For example, \cite{Nigoche-Netro:2010} by analyzing the FJR showed  that its parameters depend on the magnitude range.
%Hence, if the magnitude range is narrow the differences are negligible, but for  a wide magnitude range (as it is for cluster members), the FJR is dominated by the magnitude cut that can mask any differences caused by intrinsic physical properties of the galaxies. Concluding, FJR gives not only inaccurate but  often also biased information about the physical properties of galaxies. 

\begin{table*}[!ht]
\centering
\begin{tabular}{lrrccclll}
\hline\hline
System  &$\Delta\alpha$&$\Delta\delta$&$\epsilon$&$\theta^a$&$r_{core}$&$r_{cut}$&$\sigma_0$\\
&(\arcsec)&(\arcsec)&&&(kpc)&(\arcsec)&$(\rm km/s$) \\
\hline
 \multicolumn{3}{l}{FJ log(Evidence)=-35.1}& \multicolumn{3}{l}{${\rm RMS}_{\rm FJ} = 0.197$\arcsec} &&\\
\hline
DM 1&$8.3\pm0.3$&$-2.5\pm0.8$&$0.64\pm0.05$&$72.0\pm2.1$&$117.4\pm5.9$&[1000.0]& $884.2\pm31.9$\\
DM 2&$24.2\pm1.0$&$28.3\pm1.2$&$0.13\pm0.11$&$56.6\pm1.0$& $127.3\pm17.1$&[1000.0]&$840.2\pm37.5$ \\
DM 3&$185.9\pm0.3$&$50.1\pm0.1$&$0.46\pm0.05$&$5.2\pm0.5$& $47.3\pm0.9$&[1000.0]&$795.7\pm18.9$ \\
BCG 1&[$0.00$] &[$0.00$]&[$0.26$]&[$43.5$]&[$0.3$]&$150.0 \pm2.1$&$255.9\pm39.3$\\
BCG 2&[$24.05$] &[$29.13$] & [$0.20$]&[$37.4$]&[$0.2$]&$112.2 \pm1.8$&$201.6\pm2.4$\\
Gal A  &[$51.94$] &[$48.93$] &[$0.13$]&[$9.9$]&[$0.1$]&$60.0 \pm0.9$&$199.3\pm3.9$  \\
Gal B &[$5.23$] &[$23.01$]&[$0.10$]&$[-49.0$]&[$0.1$]&$53.3 \pm1.3$&$105.1\pm1.6$\\
$\sigma_{\rm FJ}^{\star}$&...&...&...&...&[0.1]&$48.6\pm8.5$&$119.2\pm 5.1$\\
\hline
 \multicolumn{3}{l}{FP  log(Evidence)=--34.0}& \multicolumn{3}{l}{${\rm RMS}_{\rm FP} = 0.160$\arcsec} &&\\
\hline
DM 1&$8.2\pm0.3$&$-2.3\pm0.7$&$0.56\pm0.07$&$72.4\pm2.6$&$125.8\pm7.0$&[1000.0]&$938.1\pm39.0$\\
DM 2&$24.8\pm0.8$& $27.9\pm1.0$&$0.23\pm0.09$&$55.4\pm1.1$&$133.4\pm7.6$&[1000.0]&$847.0\pm42.0$\\
DM 3&$186.0\pm0.4$&$50.1\pm0.1$&$0.44\pm0.04$&$4.8\pm0.5$&$50.0\pm3.0$&[1000.0]&$815.2\pm16.8$\\
BCG 1&[$0.00$] &[$0.00$]&[$0.26$]&[$43.5$]&[$0.3$]&$50.5 \pm12.3$&$201.8\pm32.0$\\
BCG 2&[$24.05$] &[$29.13$] & [$0.20$]&[$37.4$]&[$0.2$]&$48.8 \pm5.8$&$212.7\pm7.2$\\
Gal A  &[$51.94$] &[$48.93$] &[$0.13$]&[$9.9$]&[$0.1$]&$48.1\pm10.1$&$230.5\pm4.9$  \\
Gal B &[$5.23$] &[$23.01$]&[$0.10$]&$[-49.0$]&[$0.1$]&$48.5 \pm2.2$&$117.2\pm13.1$\\
$S_{\rm FP}$&...&...&...&...&[0.1]&$71.1\pm5.1$&$1.16 \pm 0.14^b$\\
%$\sigma_{\rm TF}^{\star}$&...&...&...&...&[0.1]&$41.1\pm1.3$&$107.1\pm 7.1$\\
\hline
\hline
 \multicolumn{3}{l}{FP log(Evidence)=-31.7}& \multicolumn{3}{l}{${\rm RMS}_{\rm FP+X} = 0.147$\arcsec} &&\\
\hline
DM 1&$9.6\pm0.4$&$-1.6\pm1.6$&$0.41\pm0.06$&$81.7\pm3.8$&$131.2\pm12.3$&[1000.0]& $918.6\pm49.6$&\\
DM 2&$21.6\pm1.2$&$25.6\pm2.1$&$0.43\pm0.07$&$64.5\pm1.8$&$108.5\pm12.5$&[1000.0]&$733.0\pm54.9$\\
DM 3&$185.6\pm0.3$&$50.1\pm0.1$&$0.30\pm0.02$&$3.5\pm0.7$&$58.9\pm1.2$&[1000.0]&$862.2\pm7.0$\\
BCG 1&[$0.00$] &[$0.00$]&[$0.26$]&[$43.5$]&[$0.3$]&$49.2\pm5.8$&$302.3\pm10.5$\\
BCG 2&[$24.05$] &[$29.13$] & [$0.20$]&[$37.4$]&[$0.2$]&$50.3\pm7.2$&$215.5\pm7.6$\\
Gal A  &[$51.94$] &[$48.93$] &[$0.13$]&[$9.9$]&[$0.1$]&$49.5\pm6.8$&$224.4\pm5.8$ \\
Gal B &[$5.23$] &[$23.01$]&[$0.10$]&$[-49.0$]&[$0.1$]&$51.4\pm7.0$&$117.5\pm13.5$\\
$S_{\rm FP}$&...&...&...&...&[0.1]&$93.1\pm59.3$&$1.19 \pm 0.76^b$\\
%$\sigma_{\rm TF}^{\star}$&...&...&...&...&[0.1]&$41.1\pm1.3$&$107.1\pm 7.1$\\
\hline
% \multicolumn{3}{l}{Xray+FP+TF log(Evidence)=-33.8}& \multicolumn{3}{l}{${\rm RMS}_{\rm FP+X} = 0.158$\arcsec} &&\\
%\hline
%DM 1&$4.97\pm0.16$  &$1.95\pm1.29$  &$0.56\pm0.02$&$85.2\pm2.1$&$105.9\pm4.9$&[1000.0]& $814.0\pm57.2$\\
%DM 2&$29.92\pm0.10$ &$25.12\pm0.88$& $0.57\pm0.01$&$54.3\pm1.6$& $117.9\pm9.1$&[1000.0]&$717.4\pm22.7$ \\
%DM 3&$189.78\pm0.26$ &$49.28\pm0.53$& $0.15\pm0.02$&$15.8\pm2.2$& $52.2\pm0.9$&[1000.0]&$788.3\pm23.3$ \\
%BCG 1&[$0.00$] &[$0.00$]&[$0.26$]&[$43.5$]&[$0.3$]&$157.5 \pm2.1$&$303.0\pm34.5$\\
%BCG 2&[$24.05$] &[$29.13$] & [$0.20$]&[$37.4$]&[$0.2$]&$134.3 \pm1.5$&$187.2\pm5.7$\\
%Gal A  &[$51.94$] &[$48.93$] &[$0.13$]&[$9.9$]&[$0.1$]&$66.3 \pm1.1$&$171.0\pm8.2$  \\
%Gal B &[$5.23$] &[$23.01$]&[$0.10$]&$[-49.0$]&[$0.1$]&$45.5 \pm1.5$&$108.7\pm5.8$\\
%$S_{\rm FP}$&...&...&...&...&[0.1]&$74.1\pm3.5$&$0.91 \pm 0.03^b$\\
%$\sigma_{\rm TF}^{\star}$&...&...&...&...&[0.1]&$47.1\pm1.0$&$101.7\pm 10.3$\\
%\hline
\end{tabular}

\caption{Modeled parameters of the three different  mass model approaches. {\bf Top} - model with Faber-Jackson scaling relation, {\bf Middle} -  model with fundamental plane scaling relation and {\bf Bottom} - our final model with explicitly included X-rays gas mass plus fundamental plane scaling relations. Values quoted within brackets were kept fixed in the optimization. The error
bars correspond to 68\% confidence levels. The location and the ellipticity of the matter clumps
associated with the cluster galaxies were kept fixed according to the light distribution. The ellipticity $\epsilon$ is the one of the mass distribution, expressed as $a^2 - b^2/a^2 + b^2$. The center is defined at $\alpha=$104.6588589 $\delta=$-55.9571863 in J2000 coordinates corresponding to the center of the first BCG.
\newline
$^a$ Position angle of the potential distribution expressed in degree, $90^{\circ}$ relative to PA.
It corresponds to the direction of the semi-minor axis of the isopotential
counted from the horizontal axis, counterclockwise.
\newline
$^b$ This is fundamental plane parameter described in Eq. 4, it is a factor S that translates  $\sigma_{\rm FP}$ into $\sigma_{\rm  dPIE}$ }
\label{modelresults}
\end{table*}

\begin{table}
\begin{center}
\begin{tabular}{lccc}
\hline\hline
System&(1) ${\rm RMS}_{\rm FJ} $&(2) ${\rm RMS}_{\rm FP} $&(3)
${\rm RMS}_{\rm Xrays}$\\
&$(\arcsec)$&$(\arcsec)$&$(\arcsec)$\\
\hline
A.1      &   0.145  & 0.193   &  0.196  \\
A.2      &   0.026  & 0.065   &  0.112  \\
A.3      &   0.119  & 0.129   &  0.125  \\
Fa.1    &   0.205  & 0.077   &  0.032  \\
Fa.2    &   0.274  & 0.078   &  0.110  \\
Fa.3    &   0.128  & 0.123   &  0.130  \\
Fb.1    &   0.183  & 0.077   &  0.110  \\
Fb.2    &   0.253  & 0.078   &  0.032  \\
Fb.3    &   0.102  & 0.123   &  0.160  \\
E.1      &   0.247  & 0.101   &  0.030  \\
E.2      &   0.254  & 0.101   &  0.030  \\
E.3      &   0.290  & 0.101   &  0.030  \\
K.1      &   0.300  & 0.230   &  0.226  \\
K.2      &   0.132  & 0.149   &  0.125  \\
K.3      &   0.215  & 0.096   &  0.104  \\
1.1      &   0.252  & 0.124   &  0.144  \\
1.2      &   0.551  &  0.124  &  0.144  \\
2.1      &   0.150  & 0.091   &  0.070  \\
2.2      &   0.085  & 0.071   &  0.046  \\
2.3      &   0.164  & 0.021   &  0.026  \\
3.1      &   0.046  & 0.245   &  0.194  \\
3.2      &   0.046  & 0.193   &  0.134  \\
3.3      &   0.046  & 0.065   &  0.318  \\
4.1      &   0.172  & 0.079   &  0.147  \\
4.2      &   0.075  & 0.079   &  0.147  \\
5.1      &   0.096  & 0.028     &  0.027  \\
5.2      &   0.056  & 0.028   &  0.027  \\
H.1    &   0.237  & 0.248   &  0.233  \\
H.2    &   0.209  & 0.375   &  0.302  \\
H.3    &   0.076  & 0.128   &  0.096  \\
I.1    &   0.367  & 0.249   &  0.132  \\
I.2    &   0.356  & 0.049   &  0.093  \\
I.3    &   0.082  & 0.284   &  0.182  \\
J.1    &   0.084  & 0.201   &  0.195  \\
J.2    &   0.084  & 0.296   &  0.201  \\
J.3    &   0.145  & 0.434   &  0.357  \\
G.1    &   0.345  & 0.036   &  0.046  \\
G.2    &   0.045  & 0.036   &  0.046  \\
6.1    &   0.054  & 0.038   &  0.045  \\
6.2    &   0.078  & 0.038   &  0.045  \\
6.3    &   0.752  & 0.038   &  0.045  \\
Total&&&\\
RMS& 0.197  &0.160&0.147\\
\hline
\end{tabular}

\caption{Goodness of fit of the three different Bullet cluster models: (1) Faber Jackson relation, (2) fundamental plane, and  (3) X-rays with fundamental plane. The RMS represents the difference  between measured position of the images and the position predicted by the model.}
\label{tab-rms}
\end{center}
\end{table}

In fact FJR is a projection of the fundamental plane (FP) relation. 
The fundamental plane \citep{Djorgovski:1987}  is a  tight correlation for elliptical galaxies between ${R_{\rm eff}}$ the effective radius, $\sigma$ the central velocity dispersion, and ${< \hspace{-3pt} I \hspace{-3pt}>_{\rm e}}$ the mean effective surface brightness, and can read:
\begin{equation}
\log R_{\rm eff}  = a \log\sigma_{\rm FP} + b \log{< \hspace{-3pt} I \hspace{-3pt}>_{\rm e}} + c,
\end{equation}
where a, b, c are free parameters of this relation.

It is understood that the FP is a consequence  of virial theorem of the dynamical equilibrium condition of elliptical galaxies. Although FP is tilted relative to the simple virial theorem prediction \citep{1997A&A...320..415B},
FP has been successfully used in various studies describing elliptical galaxies and  in strong lens mass modeling by  e.g.,  \cite{D'Aloisio:2011, Jullo:2007, Halkola:2006, Natarajan:1997}. 
The same physical motivation that led to the FJ, FP  relations apply also to galaxies within a  cluster: more luminous galaxies are more massive and rotate faster. Of course, in such a disturbed cluster like 1E 0657-56, the scatter of the relations might be larger  than in the field, nevertheless the physical motivation remains.
Therefore, in this paper we have applied FP to scale early-type cluster members.

%Unfortunately, several FP studies have shown discrepancies between the parameter values (a, b, c) depending on the samples wavelength, environment, luminosity  or redshift \citep{Jun:2008, Nigoche-Netro:2007}.
%However, there are as well studies showing the exact opposite \citep{Bernardi:2003, Treu:2001}.
%Nevertheless, to avoid problems with possible non-universality of the FP,
For the FP scaling, we employ the parameters derived by \cite{Bernardi:2003} for r-band, $a = 1.49, b = -0.75, c = 8.778$, as their galaxy sample has similar properties to the galaxy members in the Bullet cluster.
\cite{Bernardi:2003} have used a magnitude-limited sample of nearly 9000 early-type galaxies in the redshift range $0.01<z<0.3$ that was
selected from the Sloan Digital Sky Survey (SDSS) using morphological and spectral criteria. 
They concluded that FP parameters depend only little on the sample  redshifts  at $0.01<z<0.3$ and likewise they found only slight dependence on  environment. 
 Still, we apply  the redshift evolution that was found by   \cite{Bernardi:2003}, who  showed that, on average, the higher redshift galaxies are brighter, with the brightening scaling approximately as $\Delta\mu_0\approx2z$.
This directly translates to FP scaling parameters we use, the evolved parameters are now:   $a = 1.49, b = -0.75, c = 8.946$ 

We have derived $\sigma_{\rm FP}$ for each cluster member (see Eq. 3)  using the two observables  $< \hspace{-3pt} I \hspace{-3pt}>_{\rm e}$ and  $R_{\rm eff}$. 
% obtained from \textsc{SExtractor}. 
  To get a reliable estimate of the effective radius used in the fundamental plane scaling relation, we have performed a fit of the light distribution for  cluster members selected through the red sequence. We used the software {\sc Galapagos} \citep{2012MNRAS.422..449B} on the ACS/F606W Hubble image to automatically create input parameter files for {\sc galfit} \citep{2011ascl.soft04010P} on each galaxy. The light distribution is fit by a Sersic profile where we adjusted  the total flux, effective radius, ellipticity and position angle, while fixing central position and  the Sersic index to n=4 to prevent degeneracies with the effective radius. 

 Parameters $\sigma_{\rm FP}$ and $\sigma_{\rm  dPIE}$ are indeed conceptually different
quantities: $\sigma_{\rm FP}$ is the random motion of the stars and $\sigma_{\rm  dPIE}$ is normalization of the mass profile, they also differ by the radius over which they are defined, $ r_{\rm cut} >> R_{\rm eff}$ (the galaxy mass component is including both stellar mass and dark matter mass).
 Even though $\sigma_{\rm  dPIE}$ is a fiducial velocity dispersion, we wish to relate it to the measured velocity dispersion $\sigma_{\rm FP}$ of galaxies,
assuming that their profile is described by a dPIE. Therefore, we have  scaled the $\sigma_{\rm FP}$ by the factor $S$ that is optimized in the modeling process.

\begin{equation}
           \sigma_0 = S_{\rm FP}\sigma_{FP},
            \end{equation}  
 \begin{equation}                              
                           r_{\rm cut} = r_{\rm cut}^* \left(\frac{L}{L^*}\right)^{1/2},
\end{equation}   
 \begin{equation}                              
                           r_{\rm core} = r_{\rm core}^* \left(\frac{L}{L^*}\right)^{1/2},
\end{equation}  

where $L^{\star}$,  $r_{\rm core}^{\star}$  and $r_{\rm cut}^{\star}$  are, the luminosity, core radius and cut radius, the  dPIE parameters of a typical cluster galaxy \citep{Limousin:2007b}.

\subsection{X-rays}
%==============================================================================================================================
Gas in galaxy clusters represent  $\sim$10-15\% of the total mass, which can be fairly easily measured with X-rays \citep[e.g.,] []{David:1993, Neumann:2001, Vikhlinin:2006, LaRoque:2006}. However, gas is generally not included explicitly  in cluster lens modeling  \citep[with a few exceptions, e.g.,][]{Bradac:2008}. This is partly due to the fact that in relaxed clusters gas  is centered in the same region as  dark matter hence it can not be disentangled from the dark matter component. 

However, in the Bullet cluster there is a significant offset between the gas and dark matter distribution. Thus, including the gas mass as a separate component of the  mass model is  essential for realistic modeling the total mass distribution of the bullet cluster. Moreover, due to the offset, gas distribution in this cluster has been well studied \citep{Ota:2004, Markevitch:2002}.

X-ray emission in the intra-cluster gas is dominated
by thermal bremsstrahlung, and it is proportional to the line-of-sight integral
of the square of the electron density. 
Due to lack of strongly  lensed images in vicinity of the gas density centers we expect that the gas mass in the Bullet cluster provides only  an external shear to the strong lens model. 
 However,  to compare the effect of the gas mass distribution on the results, we create two fiducial models for the spatial distribution of the total (main+sub) intra-cluster gas mass. 
 centered
For our fiducial model X1, we take the spatial distribution of the total (main+sub) intra-cluster gas as a spherical  model derived from ROSAT HRI measurements
resolution of \cite{Ota:2004},
$\beta = 1.04$, $\theta_c =112.5\arcsec$, $n_{e0} = 7.2\times 10^{-3}$ cm$^{-3}$.
And we obtain a surface gas mass map by projection of the $\beta$-model gas density profile:
\begin{equation}\label{eq:xsignal}
M_{\rm gas, 2D}(r)=2\, \rho_{\rm gas,0}\, r_c \, g(\beta) \left( 1 + \frac{r^2}{r_c^2}\right)^{-3\beta/2+1/2},
\end{equation}
with $g(\beta)=1/2\,\Gamma(1/2)\Gamma(3\beta/2\!-\!1/2)/\Gamma(3\beta/2)$, $\rho_{\rm gas,0}$ the gas mass density at the center and $r_c$ the core radius.

In the  second fiducial model X2, we assume as an  approximation that X-ray emission is proportional to the square of the mass density.   We assume that this is tenable in the central region.
We take a square root of the smoothed (convolution by a Gaussian of $\sigma=2$ pixels) X-ray count map from the 500 ks Chandra ACIS-I observations
 \citep{Markevitch:2006} and we normalize it.

 The normalisation factor is computed by matching the gas mass to the one obtained by the  $\beta$ model determined from above mentioned ROSAT HRI measurements of X-ray gas mass. The total mass within  30$\arcsec$ radius of this projected model is $2.1\times 10^{13}M_{\odot}$. 

We have included the X-rays gas mass maps into our lens model using the grid technique implemented in \textsc{LENSTOOL} \citep{Jullo:2009}, and compared the results for the 2 different fiducial models.
The code iteratively  splits the $200 \arcsec\times200\arcsec$ mass map into equilateral triangles as a function of a mass threshold. At each node of this multi-scale grid a mass profile is described by a  dPIE potential  whose core radius is equal to the local grid resolution and a cut-off radius equals to three times the core radius.  We force the algorithm to stop after four levels of splitting and as a result we have  a grid cell containing $\sim$50  dPIE potentials describing the  smooth gas distribution.

%==============================================================================================================================
\subsection{Results}
%==============================================================================================================================
\begin{figure*}
\centering

\includegraphics[scale=0.35]{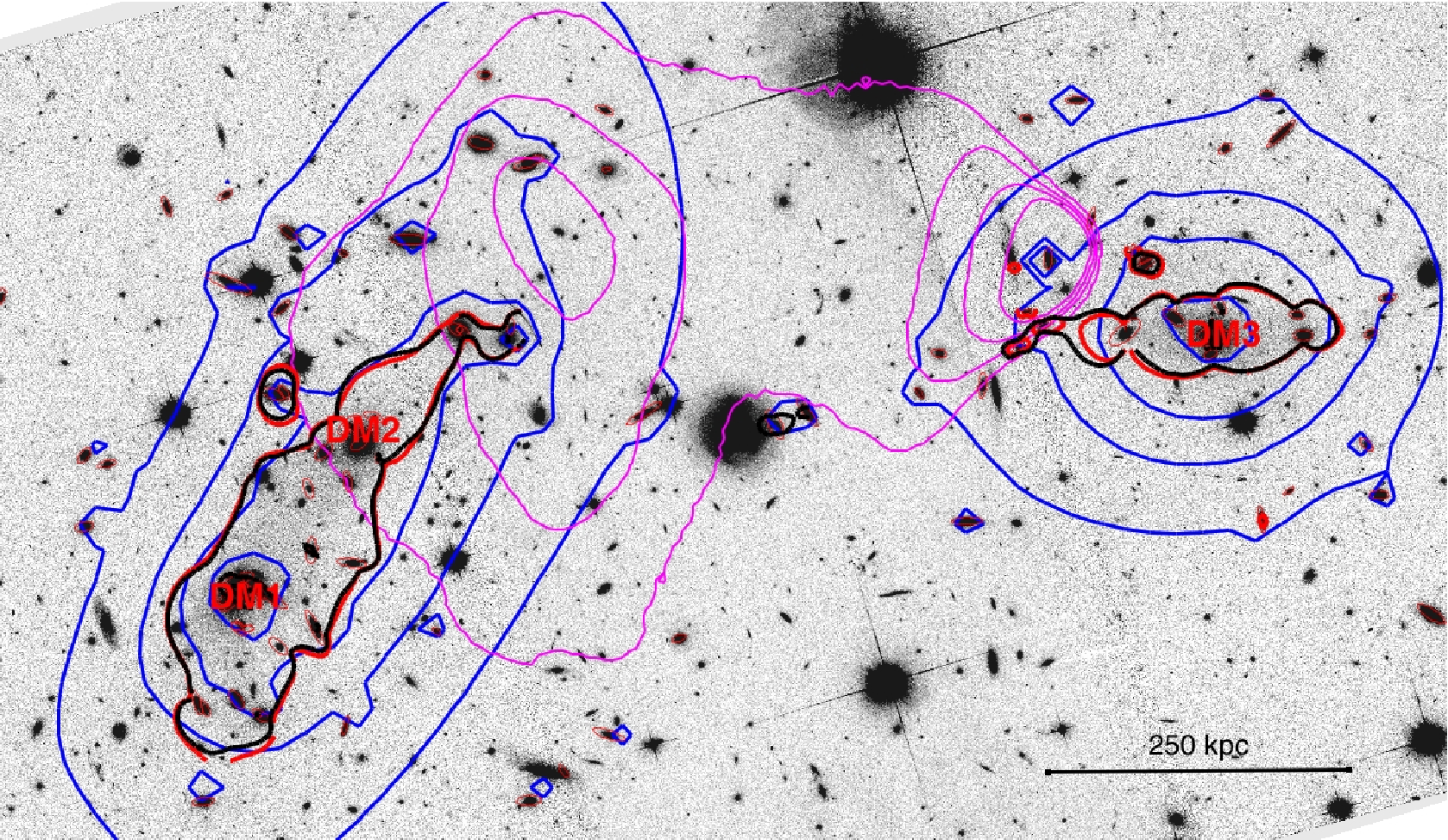}
\caption{F606W-band image of the Bullet cluster. The size of the field of view is $150\arcsec\times250\arcsec$. The blue contours show the projected
mass density.  The red line represents a critical line calculated using Faber-Jackson scaling relation to all cluster members while black line represents  the result from use of the  scaling relation, fundamental plane. The magenta lines represent the contours of the Chandra X-rays brightness map.}
\label{MassModel}
\end{figure*}

The optimized  mass model and critical lines predicted by the model at $z = 3.24$ are presented in Figure~\ref{MassModel}.
We have used the two different common scaling relation and also included the gas measured using X-rays. For our best models we find ${\rm RMS}_{\rm FJ}=0.197\arcsec$ for Faber Jackson relation, ${\rm RMS}_{\rm FP} = 0.160\arcsec$ for fundamental plane  and ${\rm RMS}_{\rm X1} = 0.147\arcsec$ and ${\rm RMS}_{\rm X2} = 0.149\arcsec$ for fundamental plane + X-rays (X1 and X2) (see Figure~\ref{MassModel}  and Table~\ref{modelresults}).

Clearly, the influence of scaling relation on overall cluster mass model is minor, the fit of the model to the data seems to be very similar  for both scalings, however it might be important for future detailed studies of galaxy clusters, for example, cluster lensing cosmography \citep{Jullo:2010, D'Aloisio:2011}.

However, the RMS of the model with explicitly included X-rays gas is  better than without  gas. The different gas mass distribution, derived with two different methods
seems not to have a significant effect on the model fitting ($\rm RSM_{\rm X1}=0.147\arcsec$  vs. $\rm RSM_{\rm X2}=0.149\arcsec$).
This is  most probably due to the flexibility of the DM model.  Apparently, the influence of gas mass is small enough that the model can compensate it by changing slightly the parameters of DM halos (position, ellipticity and size) without  losing precision of the multiple images position  reconstruction. Nevertheless, including well measured gas mass is clearly a logical choice   and does  improve the  RMS of the model. This is especially important in case of the Bullet cluster where DM and gas are spatially separated.

%==============================================================================================================================

 The Bayesian evidences \citep[see, ][]{Jullo:2007} reported in Table~\ref{modelresults} correctly summarize these observations. According to \citet{Jeffreys:1961}, the difference between two models is substantial
if $1 < \Delta \ln(E) < 2.5$, strong if $2.5 < \Delta \ln(E) < 5$ and decisive if $\Delta \ln(E) > 5$. Following
this criterion,  there is a strong evidence that model with Xray is better than those without.
%==============================================================================================================================

In Figure~\ref{MassModel} we show the F606W-band image of the Bullet cluster along with the contours generated from the projected mass map inferred from the best-fit model. 
This mass map is found to be in very good agreement with the light distribution.
 We find that the mass distribution of the Bullet cluster consists of three dark matter clumps, the main clump of the Bullet cluster is bimodal,  which is in agreement with previous models of the Bullet cluster. We find that DM1 and DM2 have high ellipticity and DM1 is comparable in mass to DM2 (see Table~\ref{modelresults}).

Furthermore, the galaxies and dark matter distributions share comparable centroid position, orientation and ellipticity. The agreement is a proof of the collisionless nature of dark matter, as suggested from the Bullet cluster by \cite{Clowe:2006}.
By integrating our two dimensional mass map, we get the total mass profile shown in Figure~\ref{MassProfile}. 
 In Figure~\ref{MassModel} we compare  also critical lines position derived by  FP relation by plotting the critical lines of the two models corresponding to z=3.24, showing good agreement of the these models.

We also compare the mass associated with the individual galaxies ($M_{\rm galax}$)  together with the 3 BCGs to the total mass ($M_{\rm tot}$) as a function of radius (see Figure~\ref{DM}). Inside radius $R< 250 \rm kpc$, we find $M_{\rm tot}=2.5\pm 0.1 \times 10^{14}M _{\odot}$,  we find also that the contribution of the galaxy halos to the total mass is $11\pm5\%$ at 250 kpc. As shown in the Figure~\ref{DM} this fraction increases towards  the center of the cluster,   similar  results were also observed in by \citet{Kneib:2003, Limousin:2007b}.

\begin{figure}
 \centering

\includegraphics[scale=0.25]{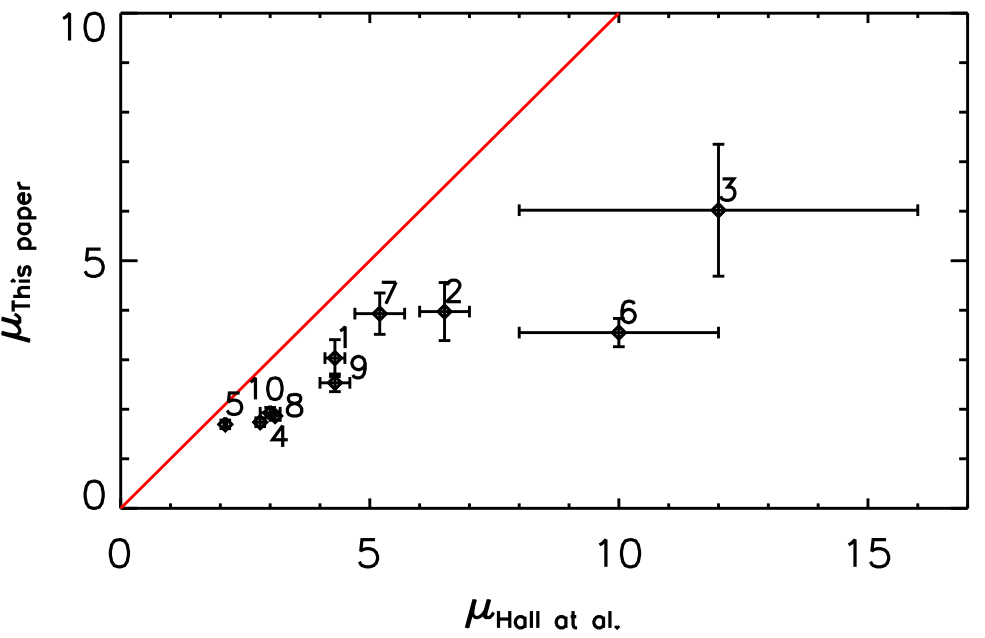}
\caption{The magnification measurements of the dropout high-z galaxies found and analyzed by \citet{Hall:2012}. We find that  average magnification of those dropouts estimated by our mass model is 43\% smaller than predicated by \citet{Hall:2012}. This is expected since high redshift critical lines of our mass model lie closer to the center of the cluster  than those of \citet{Bradac:2009}.}

\label{magnification_bullet}
\end{figure}

\begin{figure}
 \centering

\includegraphics[scale=0.55]{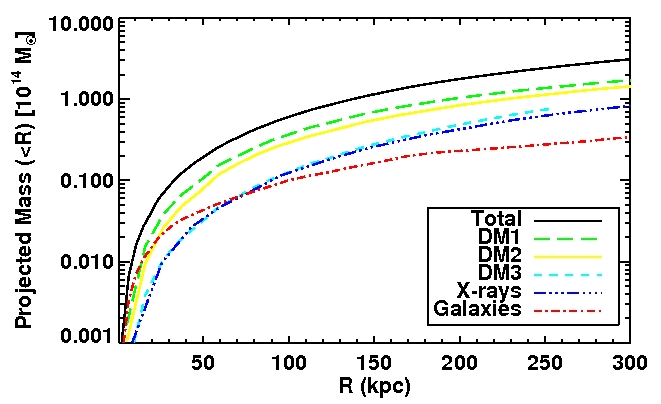}
\caption{Total projected mass as a function of aperture radius centered at BCG1 \citep[for the simplicity of the comparison with results of][]{Bradac:2006}  for different model components. The two large scale clumps, DM1 and DM2, contribute a similar amount to the mass, the X-rays gas mass measured by \citep{Markevitch:2004} is $\sim9\pm 3\%$ to the total mass at 250 kpc radius. The galaxies (including the BCG) contribute $11\pm 5\%$ within a 250 kpc radial aperture.
%% Something weird between the curves and the number given for the X-ray measurement 
%% it should be more like 18%  - need to double check the X-ray mass measurements
}
\label{MassProfile}
\end{figure}

\begin{figure}
 \centering
\includegraphics[scale=0.55]{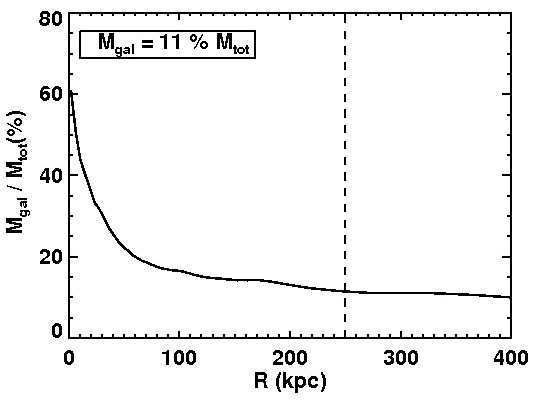}
\caption{Contribution of the galaxy component to the total mass as a function of radius (centered on the BCG 1). The vertical dotted line shows the location of the 250 kpc radius where $M_{\rm gal}=11\pm5 \%M_{\rm tot}$.}
\label{DM}
\end{figure}

As compare to previous Bullet cluster studies the main and sub clump masses estimated in this work are respectively ($11\pm4$)\% and  ($27\pm12$)\% smaller to those predicted by \citet{Bradac:2006}.  Although the difference in estimated mass is only marginally significant, the changes in the model lead to substantially different predictions for the magnification of sources near the critical lines.
 Indeed, to measure an impact of our strong lensing mass map, we have measured the magnification of the dropout high-z galaxies found and analyzed by \citet{Hall:2012}. We find that  average magnification of those dropouts estimated by our mass model is 43\% smaller than predicated by \citet{Hall:2012}, see Figure~\ref{magnification_bullet}. This is expected since high redshift critical lines of our mass model lie closer to the center of the cluster  than those of \citet{Bradac:2009}, this is specially true for southern part of the  cluster where all the droupouts are located \citep[see  Figure 1 and Table 1 at ][]{Hall:2012}.
%===============================================
%===============================================
%===============================================

We note that, our mass model along with all methods of strong-lensing mass reconstruction  have  degenerate and non orthogonal parameters.
There are numerous publications detailing  this strong lensing modeling  degeneracies  \citep[see, ][ etc]{Jullo:2007, Meneghetti:2007}.
In summary,  parameters of all lens models are clearly dependent on each other and they often compensate in order to produce
a constant enclosed mass at the images location, causing, for example,  that  the enclosed mass in the Einstein radius decreases with the model ellipticity.
The most  relevant findings for this work  was observed by \citet{Jullo:2007}  that the dPIE cut-off radius (but also the Sersic effective radius and the NFW scale radius) is one of the  less constrained parameter by strong lensing, as it lies beyond the outermost multiply imaged system. 
Moreoever, there is a severe degeneracy is between galaxy-scale subhalos  and the cluster-scale halo, especially  when no multiple images appear in the cluster.
centre.  However, they also found that the best constraints parameters were obtained in lensing configurations combining radial and tangential multiple image configurations, as it is the case of Bullet cluster. Additionally, as shown by \citet{Natarajan:1998} weak and strong lensing  provide tighter parameter constraints. Nevertheless, as it was mention before,  combining the our strong lens model with weak lensing is out of scope of this paper and will be subject of in our next work.

 Along with these degeneracies our mass model can have possible systematic error due to misidentification of multiple images and inaccuracy of their photometric redshift. 
All this can potentially make a precise model fairly inaccurate. To minimize the possibility of this  source of error we have thoroughly reviewed each strongly lensed candidate proposed by searches by checking the morphology and color agreements,  Though only spectroscopical data would give ultimate confirmation.

We estimated the redshifts of the new candidate systems using the model predictions. 
The estimated redshifts are reported in Table~\ref{mainsystem} and~\ref{subsystem}. A summary of the best-fit values inferred through the strong lensing optimization are reported in Table~~\ref{modelresults}.

  Finally, we have looked at the difference between measured $\sigma$ and $R_{\rm eff}$ of BCGs and A,B galaxy and inferred properties ($\sigma$ and $R_{\rm eff}$) of the rest cluster members.
We found that $<\sigma>=100\pm28$ ${\rm km}$ $s^{-1}$ and $<R_{\rm eff}>=56\pm14$ kpc for galaxy member, and also we find that  out of the 82 elliptical galaxies in the catalog 35 are with $\sigma< 100$ ${\rm km}$ $s^{-1}$.
 One of the two galaxies  that has been separately modeled (galaxy A) have the two properties within the average galaxy member distribution $\sigma=101\pm5$ ${\rm km}$ $s^{-1}$ and $R_{\rm eff}=43\pm2$ kpc, while the velocity dispersion of the galaxy B is bigger then average galaxy member by $79\pm37\%$ (see Figure~\ref{GalaxyMembers}).

%==============================================================================================================================

\section{Conclusions}

Due to its rare  characteristic (spatial separation of the X-ray gas and the rest of the matter), the Bullet cluster is an object of great interest for fundamental physics.
The detailed study of its total mass distribution not only brings answers about existence and nature of dark matter but also provides an exceptionally strong gravitational telescope.

In this work we have reconstructed a mass map of the galaxy cluster 1E 0657-56 using strong lensing constraints and X-rays data. 
Using deep, high-resolution optical data we have revised the previously known multiply-imaged systems and identify new ones. As a result our model is based on 14 multiply-imaged systems  with 3 spectroscopic redshifts. The model was sampled and optimized in the image plane by a Bayesian Monte Carlo Markov chain implemented in the publicly  available software \textsc{LENSTOOL}.  Our main conclusions are as follows:

1. Using the strong lensing mass reconstruction we derive a high-resolution mass
map; we get a projected, enclosed mass $M_{\rm main}(R<250 \rm kpc) =2.5 \pm 0.1 \times 10^{14 }M_{\odot}$ and $M_{\rm sub}(R<250 \rm kpc) =2.0  \pm 0.2 \times 10^{14 } M_{\odot}$.
The main and sub clump masses are respectively ($11\pm4$)\% and  ($27\pm12$)\% smaller to those predicted by \citet{Bradac:2006}. 

2. We have presented the implementation of the fundamental plane as a cluster members scaling relation and X-rays gas mass maps  into the strong lensing mass modeling. We have shown that model   with  scaling scaling relation fundamental plane  together with  explicit inclusion of X-ray gas has the best ${\rm RMS}_{\rm FP+X} = 0.147\arcsec$. The  other two models have worse, yet similar precision. The mass model without explicit X-ray gas has   ${\rm RMS}_{\rm FP} = 0.160\arcsec$  and mass model without  explicit X-ray gas and FJ as scaling relation  has   ${\rm RMS}_{\rm FJ} = 0.197\arcsec$.

3. We have found, in agreement with previous models of 1E 0657-56 that the major mass component (cluster scale-DM halos) is in spatial agreement with the  galaxies and not with the X-rays gas, which confirms the collisionless nature of dark matter.
We detect the main and sub cluster DM peak being aligned with their BCGs, both clearly offset from the location of the X-ray gas in the system.

The high precision mass map we have presented is made available to the community and can be used to exploit 1E 0657-56 as a gravitational telescope, probing the high redshift universe \citep[e.g.,][]{Bradac:2009, Kneib:2004}.

\section*{Acknowledgements}
DP acknowledges support from Agence Nationale de la Recherche 
bearing the reference ANR-09-BLAN-0234-01. 
JPK \& ML acknowledges support from CNRS. JR acknowledges support from the ERC starting grant CALENDS. JPK  acknowledge support from the ERC Advanced Grant project ``Light on the Dark'' (LIDA).
The Dark Cosmology Centre is funded by the Danish National Research Foundation.  Data presented herein were obtained as part of programs 11591, 11099, 10863,  10200 from the NASA/ESA \textit{Hubble Space Telescope}. Also partially based on European Southern Observatory program 084.B-0523 (PI: Mei). A.M. gratefully acknowledges the hospitality of the Harvard-Smithsonian Center for Astrophysics and of the NASA's Goddard Space Flight Center and he acknowledges support from Chandra grants GO2-13160A, GO2-13102A, GO4-15115X and NASA grant NNX14AI29G..

%==============================================================================================================================

\bibliographystyle{aa}
\bibliography{Bullet_2016_August}

\label{lastpage}
\end{document}